\documentclass[acmsmall,screen]{acmart}

\usepackage{amsmath,amssymb,amsfonts}
\usepackage{graphicx}
\usepackage{textcomp}
\usepackage{xcolor}
\def\BibTeX{{\rm B\kern-.05em{\sc i\kern-.025em b}\kern-.08em
    T\kern-.1667em\lower.7ex\hbox{E}\kern-.125emX}}

\usepackage[all]{nowidow}
\colorlet{RED}{red}
\usepackage{listings}
\usepackage{url}
\usepackage{makecell}
\usepackage{hyperref}
\usepackage{adjustbox}
\usepackage{xspace}
\usepackage{multirow}
\usepackage{subfloat}
\usepackage{balance}
\usepackage{tikz}
\usepackage{ctable}
\usepackage{listings}
\usepackage{color}
\usepackage[nobiblatex]{xurl}

\usepackage{multirow}
\usepackage{algorithm}

\usepackage{algpseudocode}
\usepackage{caption}

\usepackage{cleveref}
\usepackage{acronym}
\usepackage{fancybox}
\usepackage{verbatim}
\usepackage[normalem]{ulem}
\usepackage{float}
\usepackage{newfloat}
\usepackage{mdframed}

\usepackage{enumitem}
\usepackage{booktabs}
\usepackage{pifont}

\usepackage{subfigure}
\usepackage{ulem}
\usepackage[flushleft]{threeparttable}
\usepackage{booktabs}
\usepackage{tabularray}

\usepackage[utf8]{inputenc}
\usepackage{colortbl}
\usepackage[skins,most]{tcolorbox}
\tcbuselibrary{listings}

\newtcolorbox{promptcode}{
  enhanced,
  breakable,
  colback=white!0,     %
  colframe=white!0,    %
  boxrule=0pt,
  arc=0pt,
  left=0pt,right=0pt,top=0pt,bottom=0pt,
  boxsep=0pt,
  fontupper=\ttfamily\small
}

\newtcolorbox{promptbox}[1]{
  enhanced,
  colback=white,
  colframe=black!80,
  boxrule=0.8pt,
  arc=6pt,
  left=8pt,right=8pt,top=8pt,bottom=8pt,
  fonttitle=\bfseries,
  title=#1
}

\newcommand{\pfield}[1]{\noindent\textbf{#1}\ }
\definecolor{ph}{RGB}{120, 70, 200}

\newcommand{\pholder}[1]{\textcolor{ph}{\{#1\}}}

\definecolor{red}{RGB}{255,0,0}
\definecolor{green}{RGB}{18,220,168}

\usepackage{pifont}

\newcommand{\eat}[1]{}

\newlength{\MaxSizeOfLineNumbers}%
\definecolor{keywordcolor}{rgb}{0.8,0.1,0.5}
\definecolor{lightlightgray}{gray}{.96}
\definecolor{lightgray}{gray}{.925}
\definecolor{medlightgray}{gray}{0.7}
\definecolor{medgray}{gray}{0.4}
\definecolor{darkgray}{gray}{0.35}
\definecolor{nearblack}{gray}{0.15}

\crefname{component}{Component}{Components}

\newcommand{\distance}{8pt}
\definecolor{googleform-header}{HTML}{673AB7}
\definecolor{googleform-bg}{HTML}{EDE7f6}
\definecolor{googleform-frame}{RGB}{218,219,223}
\pgfkeys{/tcb/googlebox/.cd,top bar/.initial=1ex,frame arc/.initial=1mm}
\newtcolorbox{googlebox}[1][]{%
  colback=white,colbacktitle=googleform-header,colframe=googleform-frame,
  enhanced,#1,
  overlay={ \fill[googleform-header] 
    ([yshift=-\pgfkeysvalueof{/tcb/googlebox/top bar}]frame.north west) -- 
    ([yshift=-\pgfkeysvalueof{/tcb/googlebox/frame arc}]frame.north west) arc[start angle=180,end angle=90,radius=1mm]
    --
    ([xshift=-\pgfkeysvalueof{/tcb/googlebox/frame arc}]frame.north east) arc[start angle=90,end angle=0,radius=1mm]
    --
    ([yshift=-\pgfkeysvalueof{/tcb/googlebox/top bar}]frame.north east)  -- cycle;},
  boxrule=1pt,top=\pgfkeysvalueof{/tcb/googlebox/top bar}+1mm,arc=\pgfkeysvalueof{/tcb/googlebox/frame arc}
 }

\ccsdesc[500]{Software and its engineering~Software evolution}
\ccsdesc[500]{Software and its engineering~Automatic programming}

\keywords{Contextualized Code Pretraining, Code Generation, Software Evolution}

\usepackage{listings}
\usepackage{xcolor}

\usepackage{tcolorbox}
\tcbuselibrary{listings, breakable}

\newtcolorbox{mylisting}[1][]{%
  listing only,
  breakable,
  colback=white,
  colframe=black!75,
  listing options={language=Python, basicstyle=\ttfamily\small, breaklines=true},
  title=#1,
  fonttitle=\bfseries,
  enhanced,
}

\begin{document}
\title{Contextualized Code Pretraining for Code Generation}

\author{Chen Liu}
\email{cissieliu@stu.pku.edu.cn}
\affiliation{%
  \institution{Key Lab of HCST (PKU), MOE; SCS, Peking University}
  \city{Beijing}
  \country{China}
}

\author{Qingyuan Liang}
\email{liangqy@stu.pku.edu.cn}
\affiliation{%
  \institution{Key Lab of HCST (PKU), MOE; SCS, Peking University}
  \city{Beijing}
  \country{China}
}

\author{Hanwen Zhang}
\email{zhanghanwen@buaa.edu.cn}
\affiliation{%
  \institution{School of Computer Science \& Technology, Beihang University}
  \city{Beijing}
  \country{China}
}

\author{Zeyu Sun}
\email{zeyu.zys@gmail.com}
\affiliation{%
  \institution{National Key Laboratory of Space Integrated Information System, Institute of Software Chinese Academy of Sciences}
  \city{Beijing}
  \country{China}
}

\author{Yakun Zhang}
\email{zhangyk@hit.edu.cn}
\affiliation{%
  \institution{Shenzhen Key Laboratory of Internet Information Collaboration,  Harbin Institute of Technology, Shenzhen}
  \city{Shenzhen}
  \country{China}
}

\author{Lu Zhang}
\authornote{Corresponding authors: Lu Zhang.}
\email{zhanglucs@pku.edu.cn}
\affiliation{%
  \institution{Key Lab of HCST (PKU), MOE; SCS, Peking University}
  \city{Beijing}
  \country{China}
}

\renewcommand{\shortauthors}{Liu et al.}

\newcommand{\modelname}{\textsc{CallerGen}\xspace}
\newcommand{\benchmarkname}{\textsc{CallerEval}\xspace}
\newcommand{\autotoolname}{\textsc{CallerSage}\xspace}

\begin{abstract}

As code generation becomes increasingly central to improving software development efficiency, modern code models are largely trained and evaluated on code with natural-language descriptions. In real projects, developers often implement missing functions under limited project-specific artifacts, while the local call-site context is already available in the surrounding code. This usage context provides actionable cues about expected behavior, but existing models are not explicitly optimized to leverage it reliably, leading to implementations that may not integrate smoothly with surrounding usage in repository settings.

In this work, we propose contextualized code pretraining, an invocation-aware framework that integrates calling context into both the training and evaluation of code models. Using static analysis, we automatically extract large-scale caller–callee pairs from real repositories to construct pretraining tasks and benchmarks that condition generation on the calling context. We train \modelname, the first code models pretrained with invocation-aware objectives spanning multiple sizes, and evaluate them on \benchmarkname, a new benchmark featuring realistic scenarios. Experiments show that \modelname outperforms comparable-scale models and remains competitive with larger ones across two benchmarks.
Our 220M and 0.5B models achieve 16.58\% and 22.81\% pass@1, surpassing baselines on \benchmarkname. These results highlight the importance of calling context in realistic code generation.

\end{abstract}

\maketitle

\section{Introduction}

Code generation has become an essential capability for improving software development efficiency as systems grow in scale and complexity~\cite{li2022competition,svyatkovskiy2020intellicode}. As codebases expand and software engineering becomes increasingly collaborative and iterative, code generation is no longer limited to isolated function implementation; instead, it is frequently used in everyday engineering workflows such as implementing missing modules, completing APIs, and adapting code to project-specific conventions and dependencies. 
Consequently, both research and industrial practice have been moving toward code generation settings that better reflect real project development, where generated code is expected to fit into the surrounding codebase and development environment.

Meanwhile, the rapid iteration of large language models and scaling up training have substantially strengthened the base capabilities of modern code models, spanning both open models (e.g., DeepSeek-Coder and Qwen-Coder)~\cite{zhu2024deepseek,hui2024qwen2} and closed-source coding assistants (e.g., ChatGPT and Gemini)~\cite{chatgpt,team2024gemini}. Most of these models follow a relatively standard training recipe: large-scale pretraining on code or mixed with natural language using autoregressive next-token prediction, followed by instruction tuning and alignment to improve helpfulness and usability in interactive settings~\cite{transformer2017,gpt2018}. Applying these models to real software development remains challenging, and model outputs are not uniformly reliable under practical project constraints~\cite{zhang2023survey,gao2025current,liu2023your}. In parallel, repository-level code generation augments inference by retrieving similar repository contexts and organizing prompts with information from the retrieved contexts, and by incorporating execution feedback~\cite{bi2024iterative} as well as tool interactions to iteratively refine both the context and the generated code~\cite{zhang2024codeagent}. For example, RepoCoder~\cite{zhang2023repocoder} couples completion with iterative repository-context retrieval and repeatedly fetching relevant fragments from the codebase to improve completions. Note that, the retrieved context information may not be confined to information in calling contexts~\cite{liu2024graphcoder,cheng2024dataflow}. %

\begin{figure}[t]
    \centering
    \includegraphics[width=.75\columnwidth]{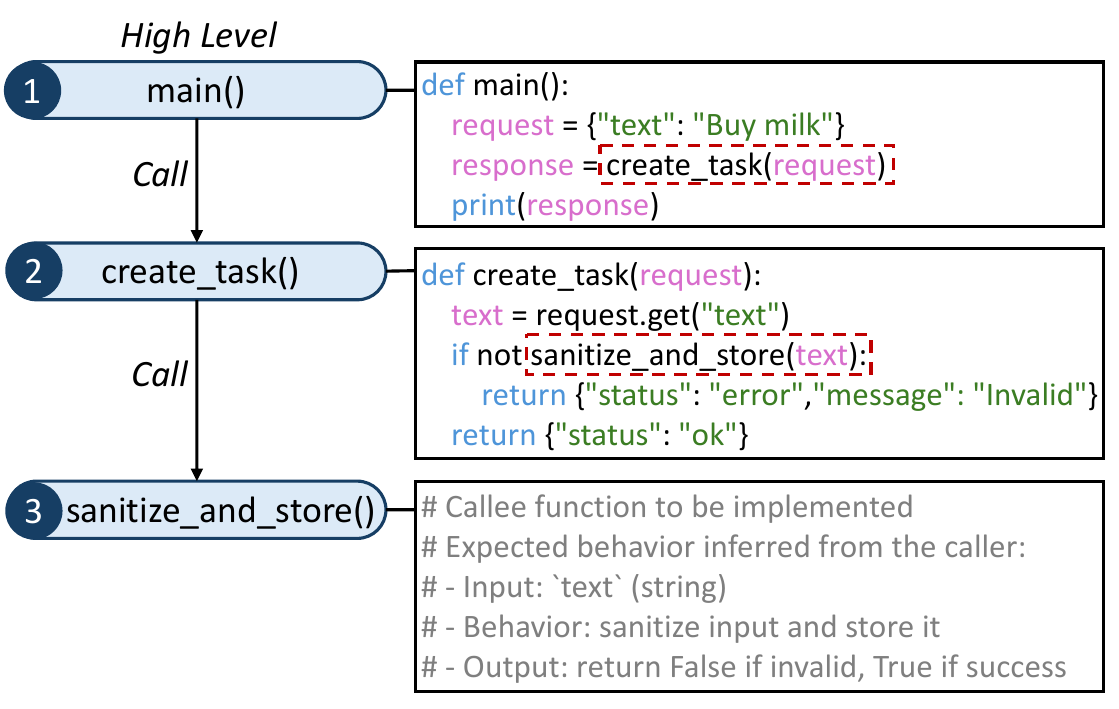}
    \caption{Web development: high-level logic first, then called functions.}
    \label{fig:sample}
\end{figure}

A limitation may arise in small or early-stage repositories, or when the functionality to be implemented is totally novel and lacks closely related exemplars in the same project. That is to say, prior approaches that attempt to augment generation with additional artifacts in the same project may be less applicable in these cases. Nevertheless, the calling context of the functionality to be implemented is often immediately available when developers start to implement the functionality~\cite{hou2024large}, and can be more readily accessible than the project artifacts surfaced by retrieval or tool-driven pipelines. In incremental development, a function is frequently introduced because it is invoked by existing logic, and its intended behavior is partially constrained by how it is called and how its results are used~\cite{roehm2012professional,latoza2020explicit}. For example, Figure~\ref{fig:sample} illustrates a typical web development scenario: a developer first writes a high-level \texttt{main()} function to process user requests. This function delegates the task to a controller-level function \texttt{create\_task}, which in turn invokes a lower-level utility \texttt{sanitize\_and\_store}, even though this callee function has not yet been implemented. Nonetheless, its behavior is often partially constrained by how it is invoked and how its outputs are consumed in the calling context.

This gap motivates us to primarily strengthen code generation with the calling context in repository scenarios where few helpful project artifacts can be surfaced and used in prompts. Our key insight is that calling contexts should be treated as an explicit conditioning signal during training. This raises three major technical challenges: \textit{acquisition}, \textit{encoding}, and \textit{evaluation}. First, \textit{acquisition} concerns how to obtain reliable caller information, as functions are invoked in diverse ways and often scattered across large codebases, making accurate data collection difficult. Second, \textit{encoding} refers to how to represent and incorporate calling contexts during training. Information about the caller is usually not directly available and is often incomplete, making it challenging to structure and utilize such context effectively for model training. Third, \textit{evaluation} refers to how to assess models under real calling contexts, requiring benchmarks with diverse caller-callee pairs that reflect real-world calling relationships.

To tackle the above three challenges, we present an invocation-aware code generation approach that explicitly incorporates calling contexts during training and evaluates models under code generation settings with calling contexts. For \textit{acquisition}, we employ an automated static-analysis process to extract invocation contexts for target functions across large Python repositories.
For encoding, we construct training examples from these collected caller-callee pairs to capture realistic usage patterns and function-level dependencies. 
We design a pretraining task that explicitly uses calling contexts as part of the input, enabling the model to learn invocation-aware generation behavior during training. To support this objective, we develop \modelname as a family of models at three different scales (60M, 220M, and 0.5B), demonstrating the approach’s effectiveness across various model capacities.
For \textit{evaluation}, we introduce a new benchmark, {\benchmarkname}, featuring diverse caller-callee pairs from real-world codebases for assessing models in code generation scenarios with calling contexts.

To evaluate \modelname, we compare it with nine open-source state-of-the-art models on two benchmarks (i.e., a public benchmark and our new invocation-aware benchmark \benchmarkname). Under settings both with and without calling contexts, \modelname consistently outperforms similarly sized baselines, narrows the gap with larger models, and in some cases outperforms even larger ones.
In scenarios with calling contexts, \modelname outperforms all baselines of similar scale. The 220M version achieves 21.22\% pass@1 on \textsc{CoderEval}, which is more than double that of CodeGen-350M (8.78\%). The 0.5B model reaches 22.81\%, outperforming Qwen2.5-Coder-32B-Instruct by nearly 2 percentage points on {\benchmarkname}. In scenarios without calling contexts, \modelname remains highly competitive, matching or outperforming baselines such as CodeGen-350M and PanGu-300M, and closely approaching much larger models. The 220M model achieves 10.00\% pass@1 from function headers, above CodeGen-350M (7.39\%) on \textsc{CoderEval}. These results demonstrate that \modelname delivers strong performance with or without calling contexts. Across all scales, invocation-aware training consistently improves results over standard models.

We summarize our contributions as follows:
\begin{itemize}[leftmargin=1.2em,topsep=0pt]

\item We formalize a new task setting \emph{code generation with calling contexts}, which emphasizes the importance of calling contexts in guiding code generation, aiming to reflect some not well suported real-world development scenarios.

\item We propose \modelname~\cite{callergen}, a pretrained model that is the first to explicitly incorporate calling contexts during training to learn invocation-aware generation behavior.

\item We construct a new benchmark \benchmarkname from real-world repositories to evaluate models under code generation scenarios with calling contexts, including realistic caller–callee invocations and a behavior-aware test protocol.

\item We conduct comprehensive experiments across public and our {\benchmarkname} benchmarks. {\modelname} achieves consistently better performance than strong baselines, and further analysis reveals how different input signals affect generation quality.

\end{itemize}

\section{Background and Motivation}
\subsection{Code Generation with Calling Contexts}

Code generation has long been a fundamental task in software engineering, enabling the automation of function synthesis, code completion, and program transformation. With the recent progress of pretrained language models in natural language processing, a growing number of models have been developed specifically for code. The representative model CodeT5~\cite{wang2021codet5} adopts an encoder-decoder architecture that pretrains on paired code and natural language samples, making it highly adaptable to diverse programming scenarios. As model capacity and data scale continue to grow, several large language models have emerged to further push the boundaries of code generation. CodeGen~\cite{nijkamp2022codegen} is trained in a decoder-only fashion on multilingual code and conversational prompts, supporting long-context generation. Similarly, PanGu-Coder~\cite{christopoulou2022pangu} adopts a multi-stage training strategy on high-quality multilingual code corpora. Driven by the rapid progress of general-purpose LLMs such as ChatGPT~\cite{chatgpt}, several foundation model families have introduced dedicated variants for code. Notably, DeepSeek~\cite{bi2024deepseek} and Qwen~\cite{qwen2.5} have released open-source models like DeepSeek-Coder~\cite{guo2024deepseekcoder,deepseekcoderv2} and Qwen-Coder~\cite{qwen2.5coder}, specifically optimized for programming tasks. These variants are available in both base and instruction-tuned formats to support downstream tasks. 

A key challenge in real-world code generation lies not in producing code that merely satisfies the functional description, but in choosing an implementation that aligns with how the function is actually used~\cite{liu2023your,wei2024magicoder,du2024evaluating}. In practice, software development often follows a top-down workflow~\cite{jorgensen2004top,mcclure2012top}: developers first write the main control logic, and then define helper functions to support it. As a result, a function's implementation is not fixed in advance, but instead shaped by its calling context. This context governs design decisions that go beyond the function’s standalone semantics. Although the function signature remains the same, its expected behavior may vary significantly depending on how it is invoked. For instance, Listing~\ref{lst:main-call} represents one possible context for usage of the user config retrieval function.

\begin{lstlisting}[caption={Invocation in main routine}, label={lst:main-call}, language=Python]
def fetch_user_config(uid):
    with open(f"/data/users/{uid}.json") as f:
        return json.load(f)
def main():
    config = fetch_user_config(uid)
\end{lstlisting}

In Listing~\ref{lst:main-call}, when the function \texttt{fetch\_user\_config(uid)} is invoked in \texttt{main()}, it is expected to simply retrieve user information from local configuration storage. 
However, the same function called from a frontend controller \texttt{render\_homepage(uid)}, as shown in Listing~\ref{lst:prod-call}, may require a different design. Here, the caller assumes that the returned object contains a stable key structure for rendering, includes language preferences, and provides a feature list that dictates UI behaviors such as displaying “BETA” or “LIVE” banners.
None of these expectations are visible in the function signature itself, they arise entirely from the caller’s assumptions about the function’s behavior and the data shape it must return.
To satisfy such contextual requirements, \texttt{fetch\_user\_config} must go beyond reading raw configuration files: it needs to infer missing attributes, normalize feature flags, and ensure consistent output structures.
This example highlights how calling contexts convey rich behavioral signals that implicitly guide a function’s design and implementation. This example illustrates how different calling contexts can lead to divergent but valid implementations of the function with the same signature, depending on what the caller needs and assumes.

Having established the context-dependent nature of function behavior, a natural question arises: do current code models learn such invocation-dependent semantics?

\begin{lstlisting}[caption={Invocation in production contexts}, label={lst:prod-call}, language=Python]
def fetch_user_config(uid):
    try:
        with open(f"/data/users/{uid}.json") as f:
            data = json.load(f)
        if uid.isascii() and uid.isalpha():
            data["language"] = "en"
        if "profile" in data:
            profile = data["profile"]
            if profile.get("type") == "tester":
                data["features"] = ["mock_mode"]
            else:
                data["features"] = ["standard"]
        else:
            data["features"] = ["default"]
    except FileNotFoundError:
        return {
            "user_id": uid,
            "language": "unknown",
            "features": [],
            "profile": {}
        }
    return data

def render_homepage(uid):
    cfg = fetch_user_config(uid)
    lang = cfg.get("language", "en")
    is_mock = "mock_mode" in cfg.get("features", [])
    banner = "BETA" if is_mock else "LIVE"
    return {"uid": uid, "lang": lang, "banner": banner}
\end{lstlisting}

{
Recent repository-level code generation research recognizes that useful evidence for implementing or completing code is often distributed across a codebase, including cross-file definitions, utilities, and usage patterns that emerge during development and integration. RepoCoder~\cite{zhang2023repocoder} organizes repository evidence through an iterative retrieval-and-generation loop, retrieving relevant fragments from the codebase and conditioning the model with information from the retrieved context to refine completion. GraphCoder~\cite{liu2024graphcoder} structures repository evidence by constructing a code-context graph and selecting context-related information in a coarse-to-fine manner, aiming to surface dependency-relevant snippets beyond surface similarity. DraCo~\cite{cheng2024dataflow} leverages extended dataflow analysis to guide retrieval augmentation so that the retrieved context better aligns with required program dependencies. CoCoGen~\cite{bi2024iterative} uses compiler feedback and static analysis to diagnose missing project-specific information and iteratively improve the retrieved context-related information and the generated code. CodeAgent~\cite{zhang2024codeagent} equips language models with repository navigation, search, and execution tools, enabling interactive evidence gathering and validation in repository settings.}

{
These approaches treat repository contexts as an inference-time resource and improve generation by retrieving and packaging repository evidence for fixed base models, via retrieval, dependency-based selection, feedback, or tool-assisted exploration.   The evidence provision objective is commonly instantiated through similarity or relevance-driven selection, which tends to emphasize exemplar generation.  That is to say, the key point of these approaches is not to teach the model to generate code under the given calling context, but to provide the trained model with information that may be useful for the code generation task. 
As a result, these approaches can be limited in two distinct ways: (1) similar code snippets may not yet exist in the repository, and (2) even when they exist, retrieval or selection may fail to surface it due to imperfect indexing, weak queries, or noisy ranking. Fundamentally speaking, because these pipelines primarily change what evidence is fed to the model, they do not directly improve the model's capability to exploit calling contexts as such signals appear in the input. } 

{
Despite variations in architecture and scale, many code models are still trained with generic language-modeling objectives on large corpora of source code~\cite{nijkamp2022codegen, guo2024deepseekcoder}. In this setup, entire files or repository fragments are fed to the model as token sequences. The objective does not explicitly distinguish call sites from function bodies, nor does it emphasize the semantic relation in which a caller imposes behavioral requirements on its callee~\cite{du2023pre}. As a result, invocation-dependent signals are not explicitly highlighted during training, and the trained models can hardly ground implementation choices on calling contexts across different usage scenarios in a reliable way~\cite{zhang2025llm,zhuo2026identifying}.}

{
In summary, existing pipelines primarily improve what evidence is provided at inference time, while prevailing pretraining objectives provide limited explicit signals for learning how calling contexts shape a callee’s behavior, leaving the utilization of invocation evidence insufficiently investigated. }

\begin{figure}
    \centering
    \includegraphics[width=.7\columnwidth]{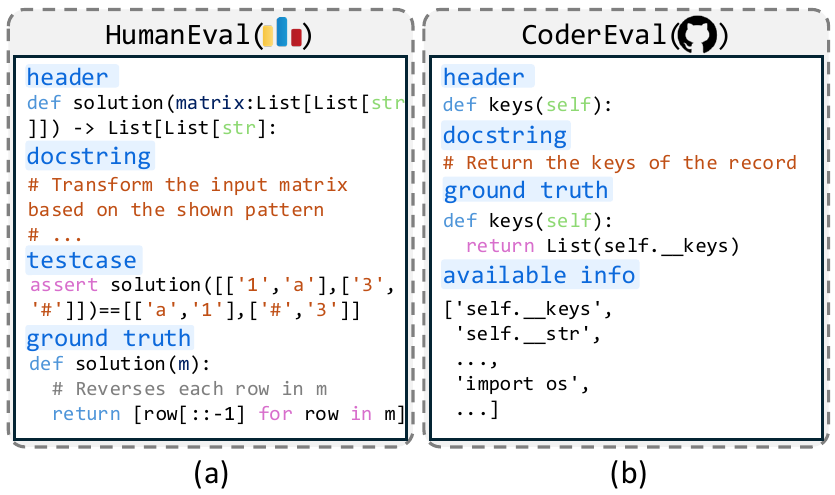}
    \caption{Comparison of \textsc{HumanEval} (a) and \textsc{CoderEval} (b).}
    \label{fig:benchmark-comparison}
\end{figure}

\subsection{Evaluation Benchmarks}
Recent advances in code generation have led to the creation of various benchmark datasets for evaluating a model’s ability to generate executable and semantically correct functions~\cite{chen2024survey,jiang2024survey}. Notably, \textsc{HumanEval}\cite{humaneval} and \textsc{MBPP}\cite{mbpp} are widely used standards. Both focus on algorithmic programming tasks, where each example provides a natural language prompt describing a well-defined problem, and the goal is to generate code that passes a set of predefined unit tests as shown in Figure~\ref{fig:benchmark-comparison} (a). While models may leverage standard or third-party libraries, these benchmarks mainly evaluate isolated function generation without requiring integration with broader project contexts or complex dependencies. As a result, they provide limited insight into how models handle the context-dependent challenges of real-world software development.

To address this limitation, \textsc{CoderEval}~\cite{yu2024codereval} collects generation tasks directly from open-source projects grounded in real-world software repositories. This design marks a significant step toward evaluating code generation under realistic project constraints.
However, each function in \textsc{CoderEval} is still presented in isolation. As shown in Figure~\ref{fig:benchmark-comparison} (b), the available information for a target function such as \texttt{keys()} typically includes its header and docstring, together with surrounding symbols like class members (e.g., \texttt{self.\_\_keys}) and file-level imports (e.g., \texttt{import os}). These elements describe the function’s outgoing dependencies (“what it needs”) but omit its actual calling contexts (“who needs it”), which convey the behavioral intent of the surrounding program. In real development, developers rarely know all internal details of a function at the beginning, but they typically have access to how the function is called, who calls it, and what behavior is expected from the calling functions.

{
Recent benchmarks further evaluate code generation in repository settings, where useful evidence is distributed across files and must be surfaced and organized. RepoBench~\cite{liu2023repobench} is designed for repository-level code auto-completion and explicitly decomposes evaluation into three interrelated tasks: retrieving cross-file context, prediction of the next line of code, and assessing end-to-end pipelines that integrate retrieval with completion. RepoExec~\cite{le2024repoexec} evaluates repository-level code generation by using execution-based checks and carefully crafted cross-file contexts to reflect project dependencies. Along the same line, CrossCodeEval~\cite{ding2023crosscodeeval} focuses on cross-file code completion and is constructed to require cross-file context for accurate completion, and it is also used to benchmark the effectiveness of cross-file retrievers. RepoCoder introduces RepoEval as a repository-level completion benchmark covering multiple completion granularities and is commonly used to evaluate retrieval-augmented repository-level completion pipelines. 
DevEval~\cite{li2024deveval} further simulates developer workflows in a working repository by pairing tasks with detailed requirements and annotated reference dependencies.}

{
Overall, existing benchmarks either evaluate standalone algorithm problems without calling contexts, or evaluate code generation with inputs that mix multiple types of repository evidence. A missing evaluation dimension is context-constrained function completion: given the designated caller-side code, the target function corresponds to an implied callee subtask, and the correctness is judged against usage requirements expressed by the calling context. This motivates a benchmark that explicitly links each target function to its real-world calling context, making the calling context a first-class element in task specification. To support faithful assessment, the benchmark should further include an execution-based evaluation component with test cases designed to reflect caller-constrained behaviors, rather than only generic functional descriptions.}

\section{\modelname}
\begin{figure*}[t]
    \centering
    \includegraphics[width=1.\linewidth]{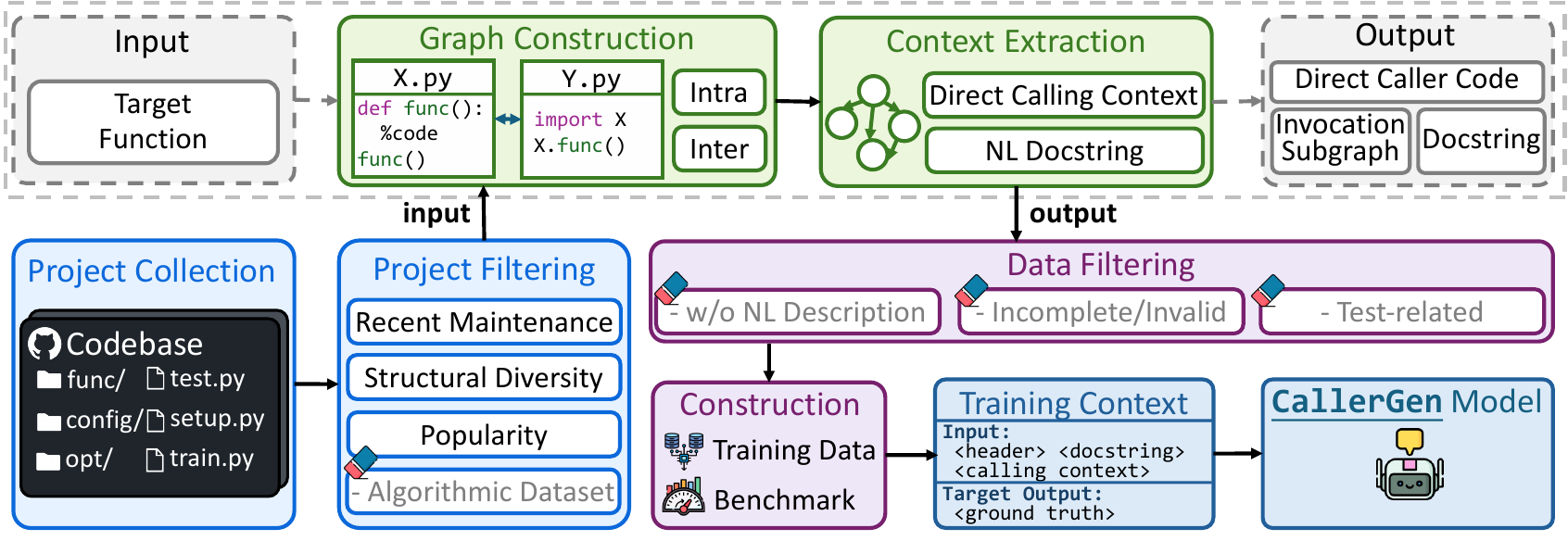}
    \caption{Overall training workflow of \modelname.}
    \label{fig:overview}
\end{figure*}

In this section, we introduce \modelname, an invocation-aware pre-trained model designed to approximate real-world software development workflows. We employ an automated static-analysis process to obtain the necessary caller-driven information from large-scale Python repositories.
\modelname is trained using a unified objective that conditions generation on the calling context. We apply this training on two types of model architectures: an encoder-decoder model (CodeT5) and a decoder-only model (Qwen2.5-Coder), allowing \modelname to generalize across diverse scenarios while supporting both architectures. The overall training workflow is illustrated in Figure~\ref{fig:overview}.

\subsection{Invocation-Aware Dataset Construction}
Obtaining reliable calling contexts from real-world projects is challenging due to scattered and diverse invocation patterns.
To systematically capture such information, we adopt an automated static analysis process that identifies function definitions and retrieves their invocation contexts across large codebases, enabling large-scale analysis of how functions are used in practice.
Based on this process, we construct a corpus of function-level samples through two stages:
(1) collecting real-world Python projects as data sources
(2) filtering and constructing the data to support both pretraining and subsequent evaluation.

\subsubsection{Automated Extraction of Invocation-Aware Contexts}
\label{sec:callgraph}
To systematically analyze the invocation cues that guide function implementation in real repositories, we employ an automated static-analysis pipeline that extracts function-level calling contexts ~\cite{salis2021pycg}. Since functions are typically written in response to upstream usage, we use static analysis to automatically recover their caller-driven contexts. The extraction proceeds by locating each target function in the repository, resolving its invocation relationships, and retrieving the caller-side code that captures usage contexts.

The process begins by parsing each Python source file into an abstract syntax tree (AST) using the built-in ast module. We recursively traverse the tree to collect function and class declarations, their fully qualified names, and all import statements (e.g., import x, from x import y as z). Alias mappings introduced by these imports (e.g., np $\rightarrow$ numpy, bar $\rightarrow$ utils.foo) are recorded to support accurate resolution of cross-module references. Global variables defined at the top level are also included, as they may influence function behavior or participate in call chains.  All extracted symbols are stored in per-file symbol tables, which later support name resolution in both local and cross-module references during call graph construction.

After extracting symbol information, we analyze the body of each function to identify all function and method calls. For every call expression (i.e., Call node in the AST), we determine which function is being invoked by consulting three sources: local definitions within the same file, imported functions recorded in the symbol table, and class methods identified through inheritance analysis. From these resolved relationships, a function-level call graph is constructed, where each node represents a function and each edge denotes an invocation. We maintain bidirectional edges “calls” and “calledby”, to enable both forward and backward traversal. Both intra-file and inter-file invocations are captured to ensure comprehensive coverage of repository-level dependencies.

After the call graph is built, we extract all direct callers of the target function and treat their complete function bodies as the invocation-aware context. This one-hop scope reflects concrete and immediate usage scenarios, such as how return values are processed, what error conditions are handled, or how inputs are prepared. While avoiding unnecessary complexity from higher-level call chains that provide weaker behavioral cues.

In parallel, we extract the natural language docstring of the target function by locating the string expression immediately following its AST node. These descriptions are often written either before implementation, as a way to articulate the intended behavior during development. As such, they reflect a semantic intent grounded in human reasoning rather than program structure. Not all functions include such descriptions, the docstring can serve as an additional input signal that complements the invocation-aware code context when available. 

\subsubsection{Real-World Project Collection}
\label{sec:realworldpro}
Applying the automated static analysis process to a wide range of real-world Python repositories, we can obtain realistic caller-driven scenarios and diverse function usages.
Compared to platforms centered on programming exercises, such as algorithmic problem sets, GitHub repositories provide access to real projects with meaningful module organization and both intra-file and inter-file dependencies. These projects reflect how functions are defined and invoked in practice, offering a more representative foundation for studying invocation-aware code generation tasks.

Specifically, we focus on Python projects that are widely used for many domains, such as machine learning, web development, and automation scripting. Its dynamic semantics and concise syntax facilitate static analysis. Compared to other languages, Python exposes structural dependencies more transparently, enabling more reliable extraction of file scopes, module-level imports and function-level interactions.

Given the scale and diversity of available repositories, we do not include all candidates. Many projects consist of minimal examples, tutorial code, or highly synthetic patterns that lack structural depth. Including such data may introduce noise or bias into model training. To ensure relevance and quality, we select 800 Python repositories by applying the following filtering criteria:

\begin{itemize}[leftmargin=1.2em,topsep=0pt]
    \item \textbf{Popularity}: Repositories with at least 100 stars, indicating community engagement and non-trivial usage.
    \item \textbf{Recent Maintenance}: We prioritize repositories with evidence of ongoing maintenance (e.g., recent commits) to reduce the inclusion of outdated or deprecated code.
    \item \textbf{Structural Diversity}: Projects with multiple files and modules, rather than a single script or a flat structure.
    \item \textbf{Exclusion of Algorithmic Datasets}: We remove repositories that primarily consist of competitive programming problems or academic algorithm implementations. These repositories tend to contain isolated functions with minimal contextual linkage, lacking the complexity found in production codebases.
\end{itemize}

\subsubsection{Data Filtering and Construction.} 
\label{sec:datapre}
Based on the 800 repositories collected in the previous step, we automatically extract a large number of \emph{target functions} together with their invocation contexts (real callers) and optional natural-language descriptions. 
We remove functions that lack descriptions and filter out those that are incomplete or syntactically invalid to ensure code quality and executability. To further improve training data quality and prevent potential leakage, we eliminate functions identified as test-related, such as those located in testing directories (e.g., \texttt{*/tests/*}, \texttt{test\_*}) or containing dense assertion patterns, so that no unit-test or evaluation code is included, and no test functions are ever treated as callers during extraction. These filtering steps ensure that the remaining functions represent genuine implementation code with high executability and minimal noise. After preprocessing, we retain 235{,}460 unique target functions. Each data instance in our corpus corresponds to a single function paired with its extracted calling contexts and natural-language description. 
{
If a target function is invoked by multiple distinct callers, we treat each caller--callee pair as a separate training instance, i.e., one caller context per instance, rather than concatenating multiple call sites into a single example. Additional callers (when present) will form additional instances. This fixed-budget formulation preserves invocation-level supervision while avoiding prompt-length and truncation artifacts, yielding a consistent serialization format.
Importantly, this does not discard multi-caller supervision: if a function has $m$ callers, it contributes $m$ one-caller instances to the training corpus, exposing the model to diverse usages in an implicit but scalable manner.
To support unbiased benchmarking, we enforce repository-level non-overlap so that code data used for model pretraining does not contribute to subsequent benchmark construction. We describe the benchmark construction protocol in Section~\ref{sec:benchmark}.}

\begin{table}[t]
\centering
\caption{Token length statistics for training instances.}
\label{tab:token-statistics}
\begin{adjustbox}{max width=1.\columnwidth}
\begin{tabular}{lccc}
\toprule
\textbf{Percentile} & \textbf{Task Length} & \textbf{Target Code Length} & \textbf{Total Length} \\
\midrule
Mean     & 419.7   & 176.3   & 596.0 \\
Median   & 244.0   & 80.0    & 371.0 \\
90\%     & 879.0   & 393.0   & 1233.0 \\
95\%     & 1261.0  & 652.0   & 1787.0 \\
99\%     & 2770.0  & 1452.0  & 3306.0 \\
\bottomrule
\end{tabular}
\end{adjustbox}
\end{table}

\subsection{Pretraining Tasks}
\label{sec:pretraining-task}
With samples constructed as described in Section~\ref{sec:datapre}, we define an invocation-aware training that reflects how functions are written and used in real codebases. Rather than relying solely on natural language inputs, we aim to model function generation as a contextual prediction task guided by upstream usage.

Given each training instance, the model is conditioned on three inputs: (1) the function signature (i.e., header); (2) the code snippet of a direct caller function; (3) the original docstring of the target function. The learning objective of this model is to generate the full body of the target function based on these contextual cues. This formulation allows the model to integrate structural invocation signals with semantic intent, leveraging both code and language guidance to predict function behavior. 
For each training instance, we serialize the invocation-aware context by concatenating all relevant components into a single text sequence. This sequence is then tokenized using standard byte pair encoding (BPE), ensuring compatibility with both encoder-decoder and decoder-only architectures.
Specifically, the sequence is composed of:
\begin{equation}
x = \texttt{<func>}, h, \texttt{<calledby>}, c, \texttt{<docstring>}, d
\end{equation}

\begin{figure}[t]
    \centering
    \includegraphics[width=.7\columnwidth]{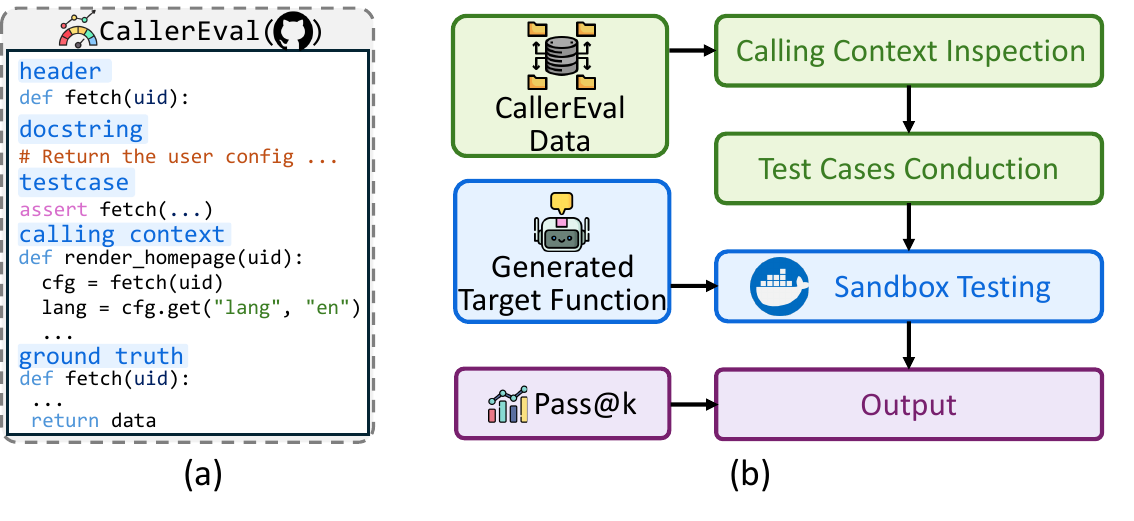}
    \vspace{-1em}
    \caption{(a) A sample of \benchmarkname. (b) The end-to-end evaluation process of \benchmarkname.}
    \label{fig:benchmark-pipeline}
\end{figure}

Where $h = \{h_1, \cdots, h_n\}$ represents tokens from the function header, $c = \{c_1, \cdots, c_m\}$ is the set of tokens from the code snippet of caller functions, and $d = \{d_1, \cdots, d_k\}$ denotes tokens from the natural language docstring. The markers \texttt{<func>}, \texttt{<calledby>}, and \texttt{<docstring>} serve as delimiters to highlight semantic boundaries.

Formally, the generation task is defined as learning a conditional language model $P(y \mid x)$, where $x$ is the context sequence described above, and $y$ is the target function’s full implementation. The model is trained to maximize the log-likelihood of the output sequence.

\subsection{Model Architecture}

\modelname adopts invocation-aware training on two different model architectures to demonstrate the generality of our approach. For the encoder-decoder architecture CodeT5, we pretrain on both the 60M version (CodeT5-small) and the 220M version (CodeT5-base) by incorporating calling contexts into the training objective. For larger-scale, we apply continued pretraining on a strong open-source decoder-only base model (i.e., Qwen2.5-Coder-0.5B). This design allows each model to preserve general coding capabilities while adapting to caller-driven code generation.

\section{\benchmarkname}
\label{sec:benchmark}
This section details the construction of \textsc{CallerEval}, a benchmark designed to evaluate code generation under realistic caller-driven scenarios. We describe the design principles, test case construction process, automated evaluation protocol, and dataset statistics.

\subsection{Design Principles}
{
A primary design principle of \benchmarkname is strict independence from the data used for model pretraining. We enforce repository-level non-overlap between the pretraining corpus and benchmark construction so that benchmark tasks are drawn only from repositories that do not contribute code to model training. This protocol ensures that the evaluation measures generalization to unseen projects rather than memorization within the same codebase.}

{
We construct \benchmarkname as a functional benchmark, where each task is evaluated by executing code and running tests under a controlled environment. As a result, a task is eligible for inclusion only when its dependencies and test harness can be reliably established and reproduced.    We intentionally keep \benchmarkname at a manageable scale and prioritize reliable, reproducible functional evaluation over sheer scale. Our benchmark aims to (1) preserve real caller–callee semantics, (2) enable function-level execution with minimal dependency setup, and (3) support reproducible, automated testing at scale. When multiple call sites are available, we use them to guide test construction so that the resulting suite exercises behavioral constraints implied by diverse real usages, approximating a repository-level regression contract rather than fitting a single invocation.
These principles guide the design of the benchmark and the workflow described below. Concretely, \benchmarkname contains 263 function-level generation tasks. As shown in Figure~\ref{fig:benchmark-pipeline}(a), each task comprises a target function paired with one or more of its real callers, a natural-language description and an accompanying behavioral test suite. 
}

\subsection{Test Case Construction}
To faithfully evaluate whether generated functions behave as expected in invocation-aware scenarios, we design lightweight behavioral test suites guided by their real calling contexts. The functions in \textsc{CallerEval} are extracted from large open-source repositories based on caller–callee relations. While some repositories include partial unit tests, most target functions lack directly reusable cases, or their tests depend on project-level fixtures. To enable consistent function-level evaluation, we therefore integrate two complementary strategies: (1) reusing and converting existing tests whenever feasible, and (2) reconstructing lightweight behavioral tests when direct reuse is infeasible.

For functions with reusable tests, we automatically parse the original test files, 
identify cases that call the target function through static call-graph analysis, and extract the relevant statements. We then normalize these test fragments into self-contained drivers and convert them into standard \texttt{main()}-style scripts that can execute independently. 

When no reusable tests exist, we supplement the suite by designing targeted scenarios derived from the function’s real calling contexts. This reconstruction follows a three-stage process that centers on constructing usage patterns, deriving the behavior sketch, and instantiating them into test scenarios. 
Specifically, for a target function $f$, let $\mathcal{C}_f = \{c_1,\dots,c_n\}$ denotes all its call sites in the repository. Each call site induces a set of observable semantic requirements on~$f$. Different patterns may impose overlapping semantic properties (e.g., both requiring the result to be a dictionary), but represent different ways the result is actually used in the code. We denote this requirement set by $R(c_i)$ for each $c_i$.
To organize these call sites, we group them into a small set of \emph{usage patterns} $\mathcal{U}_f = \{U_1,\dots, U_k\}$, where each pattern collects callers that rely on~$f$ in a similar and mutually compatible way. Within a pattern, callers tend to interact with~$f$'s output through the same kinds of operations, for instance, accessing the same field of a returned dictionary (e.g., \texttt{cfg["language"]}), invoking the same method on a returned object, or handling the same exception type. Formally, each pattern is associated with the requirements shared by all its members, which yields all behavioral constraints that are exercised by callers in that pattern:
\[
    R(U_j) = \bigcup_{c \in U_j} R(c),
\]

We then combine these per-pattern requirements into a lightweight \emph{behavior sketch} for the function:
\[
    B(f) = \bigcup_{j=1}^{k} R(U_j).
\]

The behavior sketch provides a checklist of properties that $f$ must satisfy to remain compatible with the ways it is used across the caller snippets. This sketch then guides the construction of test scenarios. Each scenario instantiates one or more items from $B(f)$ using concrete inputs and assertions, and the suite as a whole is required to exercise all usage patterns at least once. This process allows interaction modes (captured by usage patterns) and semantic requirements (captured in the behavior sketch) to be treated cleanly and systematically during test construction.

To ensure consistency and completeness, each test suite $T_f$ must satisfy three coverage criteria derived from the usage patterns and behavior sketch:
\begin{itemize}
    \item \textbf{C1 (Usage-pattern coverage).}
    Each pattern $U_j \in \mathcal{U}_f$ must be exercised by at least one scenario in $T_f$, ensuring that all interaction modes observed in real callers are represented.

    \item \textbf{C2 (Behavior-requirement coverage).}
    Let $B(f)$ be the behavior sketch aggregated from all usage patterns. Every semantic requirement in $B(f)$ must be covered by at least one assertion in $T_f$. Because scenarios may verify several requirements together, the suite can cover all elements of $B(f)$ without needing a separate scenario for each individual requirement.

    \item \textbf{C3 (Evidence-grounded assertions).}
    Every assertion in $T_f$ could be justified by explicit evidence in the original project, either from caller code or from function-level documentation (e.g., docstrings or type hints).  
\end{itemize}

Although a function may have many call sites, callers within the same project typically reuse the function in only a few consistent interaction modes.   As a result, the set of usage patterns $\mathcal{U}_f$ is small, and the unified behavior sketch $B(f)$ contains a limited number of distinct semantic requirements.   A small number of scenarios is sufficient: individual scenarios can instantiate multiple items from $B(f)$, and different scenarios can emphasize different interaction modes from $\mathcal{U}_f$. We cap each suite at five scenarios, which provides enough capacity to cover all observed patterns as well as representative normal, boundary, and error-handling behaviors, while keeping each scenario concise and interpretable.

To ensure reliability, three researchers with software engineering backgrounds independently constructed and cross-validated all test drivers, resolving inconsistencies through discussion and re-verification. This hybrid process, combining partial reuse, automated extraction, targeted reconstruction, and multi-stage validation, ensures that every function in \textsc{CallerEval} is executable, semantically meaningful, and evaluated under a consistent, reproducible, and caller-driven protocol.

\subsection{Evaluation Process}
To support automated evaluation at scale, we implement a containerized testing platform based on Docker. The platform executes each generated function inside a sandboxed runtime environment, replacing the original implementation and invoking the function with the constructed test inputs. The outputs are then compared against the expected values derived from the behavioral test specification.

Our framework supports evaluating \texttt{pass@k} performance, where multiple generations per function can be tested. A generation is considered successful if it satisfies all test cases associated with the target function. This setup allows us to assess not only whether a model can generate one correct implementation, but whether it consistently produces valid variants that meet the behavioral contract implied by the calling code. Figure~\ref{fig:benchmark-pipeline} (b) illustrates this end-to-end evaluation process.

\subsection{Dataset Statistics}
\label{sec:stats}
\benchmarkname encompasses a wide range of real-world software projects drawn from diverse application domains, spanning data analytics, scientific computing, machine learning, web and backend systems, command-line utilities, and other general-purpose libraries, ensuring variation in both function granularity and dependency complexity. Such diversity allows us to evaluate whether models can generalize across heterogeneous invocation patterns rather than overfitting to domain-specific conventions.

To verify that the benchmark faithfully reflects real development practice, we first analyzed the large corpus of GitHub repositories collected in Section~\ref{sec:realworldpro}. Nearly all functions have at least one caller, about 60\% have two or more, and over 30\% have three or more, averaging roughly five callers per function, indicating that functions in real-world projects are rarely isolated. Data in \benchmarkname exhibits a highly similar distribution: every target function has at least one caller (100.0\%), 62.4\% have two or more, and 45.6\% have three or more, with an average of 4.22 callers per function. This alignment confirms that \benchmarkname captures the caller-driven structures commonly observed in practical software ecosystems, providing a realistic basis for evaluating invocation-aware code generation.

Caller snippets range from concise single-line invocations to multi-step logic involving normalization, error handling, and chained API operations. Roughly one quarter of targets appear within explicit \texttt{try--except} constructs, and over ninety percent participate in chained or method-style calls.  These rich invocation patterns impose concrete behavioral and interface constraints on generation models, highlighting \benchmarkname’s realism and its suitability for assessing invocation-aware code generation.

In terms of function complexity, most target functions are of moderate length, typically within twenty lines of code, while some reach around thirty to forty lines. The number of parameters also varies, with many functions taking one or two arguments and others involving three or more.  This diversity design introduces a natural hierarchy of difficulty, allowing evaluation across both simple and more structurally complex targets.

As summarized in Table~\ref{tab:dataset-summary}, \benchmarkname exhibits realistic invocation patterns and balanced task complexity, providing a representative and analytically tractable basis for evaluating invocation-aware code generation.
\begin{table}[t]
\centering
\caption{Statistical summary of \textsc{CallerEval}. }
\label{tab:dataset-summary}
\setlength{\tabcolsep}{6pt}
\begin{tabular}{lcc}
\toprule
\textbf{Aspect} & \textbf{Statistics (Mean)} \\
\midrule
Invocation count (Callers per target)  & 4.22 \\
Function length (Code Lines) & 19.8 \\
Parameters (Num)       & 1.78 \\
Caller context (Lines per caller)    & 21.33 \\
\bottomrule
\end{tabular}
\end{table}

\section{Experiment Setup}
\label{sec:setup}

To evaluate our invocation-aware training under realistic development scenarios, we focus on two representative input settings.

The first is the caller-available setting, where both the function header and its actual calling context are available as input (optionally with natural language descriptions). This reflects situations in real development where the calling context can guide function implementation, which is the primary use case for our approach.

The second is the non-caller setting, which includes only the function header or the header with a natural language description. This setting is common in prior benchmarks where the calling context is not available.

We choose these two settings because they cover the typical situations faced by developers, either having access to calling context or not. Evaluating both allows us to test the effectiveness and generalizability of our approach. These scenarios motivate the following research questions.

\noindent\textbf{RQ1: Can \modelname better support caller-available development scenarios?}
To answer this, we evaluate \modelname on inputs that include the function header and its real-world callers, under two configurations: with and without natural language descriptions. Specifically, we compare model performance on two different benchmarks under this development-aligned setting.

\noindent\textbf{RQ2: Can \modelname remain effective in non-caller settings?}
To tackle this question, we test the model under inputs that provide only the function header, or the header along with a natural language description. 
Without calling contexts, common in prior work, we compare \modelname with other models across both benchmarks. 

\noindent\textbf{RQ3: Does our invocation-aware training consistently benefit different base models?}
To address this question, we compare our models trained in two different ways on CodeT5 (60M/220M) and Qwen2.5-Coder (0.5B) to investigate the relative gains across model families.

\noindent\textbf{RQ4: What aspects of caller context matter?}
To analyze this problem, we disentangle the contribution of different caller cues and study how performance changes as we aggregate multiple callers, using controls to separate semantic evidence from length and surface forms.

\subsection{Downstream Tasks}
\label{sec:DownstreamTasks}

Our evaluation focuses on realistic, context-rich code generation scenarios, where the goal is to produce functions that not only pass tests but also integrate correctly with their real-world callers.
Unlike benchmarks that emphasize solving stand-alone algorithmic problems from natural-language prompts, our study targets invocation-aware generation in practical software codebases.
Accordingly, we primarily evaluate on our proposed dataset \textsc{CallerEval}, and additionally verify generalizability on the public benchmark \textsc{CoderEval}~\cite{yu2024codereval}.

\textbf{CallerEval.}
The first dataset \textsc{CallerEval} is designed to capture the caller-available nature of real-world software development.
Each of its 263 tasks consists of a target function paired with one or more actual caller functions extracted from open-source projects.
This pairing explicitly exposes realistic calling contexts that describe how each target function is invoked and used within its surrounding code. As analyzed in Section~\ref{sec:stats},
each target function in \textsc{CallerEval} is associated with realistic caller contexts
(averaging 4.22 callers per target),
closely matching the dependency patterns observed in large-scale open-source projects.
This design enables more faithful evaluation of invocation-aware generation.

\textbf{CoderEval.}
The second dataset \textsc{CoderEval}~\cite{yu2024codereval} is a widely-used benchmark for function-level code generation. We select the Python subset of \textsc{CoderEval}, which includes 230 function generation tasks. Each task in \textsc{CoderEval} provides the method name, natural language description but lacks information of the actual caller functions.
To enable a fair comparison without altering the benchmark itself, we automatically supplemented each task with representative caller snippets obtained or reconstructed from its original repository, which drawn from 43 repositories. Due to the time-varying accessibility of third-party repositories (e.g., deleted or history changed), 9 repositories were not consistently retrievable during our analysis;  the remaining 34 indexed repositories range in size from 42.11KB to 90.05MB and contain between 5 and 1325 Python files. The benchmark tasks are not uniformly distributed across repositories; in particular, up to 28 targets originate from the same repository. When real call sites were available, we extracted them via static dependency analysis. When a target had no observable callers, we synthesized a minimal, type-consistent invocation based on its signature and surrounding context to emulate a realistic usage entry point.

Across the 230 Python tasks in \textsc{CoderEval}, static dependency analysis identifies at least one \emph{real} call site for 164 targets (71.3\%), yielding 1312 distinct call-site hits in total; the remaining 66 targets (28.7\%) have no observable callers in the retrieved repositories and therefore use a synthesized minimal invocation as the context entry point. Among the real-call subset, caller availability is sparse. The median target has two recovered call-site hits, and for 107 targets (46.5\%) the number of observable real callers is fewer than five, indicating that many task are only weakly connected through explicit call sites in their projects.
These snippets are used only as context inputs during model inference; all task definitions, unit tests, and evaluation scripts of \textsc{CoderEval} remain unchanged. This ensures full compatibility and comparability with prior work, while allowing us to analyze whether invocation-aware training improves the model's capacity to handle real-world calling contexts.

To ensure the validity of our evaluation, we carefully prevent any potential data leakage between training and test datasets. First, we confirm that no function or repository in \textsc{CoderEval} overlaps with our training corpus, validated through function-level and repository-level cross-checking to ensure that neither target functions nor their caller contexts appear in the training data.
Second, as described in Section~\ref{sec:datapre}, \textsc{CallerEval} was constructed with a strict split that prevents any overlap between training and evaluation instances.
We explicitly excluded all test functions (e.g., unit tests or benchmark cases) from being used as callers, preventing any potential leakage of evaluation signals.
Together, these measures ensure the fairness and validity of our evaluation, providing an unbiased assessment of each model’s generalization ability.

\subsection{Metrics}
\textbf{Pass@K.}
We primarily evaluate functional correctness using \textbf{Pass@K}, which directly aligns with the objective of the code generation task: producing an executable implementation that satisfies the specification as captured by tests. For each generation task, a solution is considered correct if \textit{at least one} of the top-$K$ outputs passes all associated test cases. Pass@K thus measures the proportion of tasks for which a valid solution appears within the top $K$ generated candidates. We report Pass@K for $K \in \{1, 5\}$. To ensure statistically reliable estimates under limited sampling, we adopt the unbiased estimation procedure established in prior work~\cite{chen2024survey,humaneval,yu2024codereval}, which accurately reflects the probability of observing correct outputs among the sampled completions.

{Given that ground-truth functions are available in our benchmarks, we additionally report \textbf{CodeBLEU} and \textbf{ROUGE-L} as additional static metrics to characterize similarity between the generated code and the ground-truth implementation. Since code often admits multiple semantically equivalent implementations, reference-based similarity may vary even among functionally correct solutions. Therefore, we treat Pass@K as our primary metric for correctness, and report CodeBLEU/ROUGE-L as complementary signals that provide additional perspectives on lexical and structural proximity to the ground-truth implementation. }

{
\textbf{CodeBLEU.}
CodeBLEU~\cite{ren2009method} is a code-oriented similarity metric intended to capture not only lexical overlap but also \emph{structure- and dependency-level} similarity between a generated program and a reference implementation. It combines (i) standard BLEU, (ii) keyword-weighted $n$-gram match, (iii) AST-based syntactic match, and (iv) data-flow based semantic match via a weighted sum. In our experiments, we adopt the widely used a public implementation with its default weights (i.e., $\alpha,\beta,\gamma,\delta = 0.25, 0.25, 0.25, 0.25$) 
, to make our reported scores directly comparable to those commonly reported in the literature.
}

{
\textbf{ROUGE-L.}
ROUGE-L~\cite{lin2004rouge} is a sequence-level overlap metric designed to measure how well a generated sequence matches a reference sequence, which is based on the length of the \emph{longest common subsequence} (LCS) between the generated and reference token sequences. We report the corresponding F-measure (denoted as ROUGE-L$_{F1}$), where precision and recall are computed from the LCS length over tokenized code and combined as $F1 = \frac{2PR}{P+R}$.}

\subsection{Compared Baseline Models}
To evaluate the effectiveness of our approach, we compare it against a diverse set of baseline models spanning both encoder-decoder and decoder-only architectures at various scales. Due to resource constraints, we focus on evaluating models at a range of scales, which limits the number of configurations tested. However, despite these limitations, we ensure the reliability of our experiments by carefully selecting a representative set of baseline models and maintaining consistency in training protocols and evaluation metrics. 

{
For decoder-only models, we first evaluate against eleven widely-used baselines, including \textbf{CodeGen-Multi} (350M and 2B) and \textbf{PanGu-Coder} (300M), which are commonly adopted for function-level code generation. Note that, \textbf{CallerGen-qwen (0.5B)} is trained from \textbf{Qwen2.5-Coder-0.5B} using the invocation-aware objective, enabling direct comparison with its original pretrained counterpart. To ensure consistency within the same model family, the \textbf{Qwen2.5-Coder-Instruct} series (0.5B to 32B) is included as a reference baseline. In addition, \textbf{DeepSeek-Coder-Base} (1.3B and 6.7B) is considered for its recently demonstrated strong performance across various code generation benchmarks. Finally, we include \textbf{Codex} (GPT-5.2) as a strong proprietary code model reference. All models are evaluated using prompt templates adapted to their expected input formats.
For encoder-decoder models, we include a trained context-free version on \textbf{CodeT5-small} (60M) and \textbf{CodeT5-base} (220M) to clarify the role of calling context. }

{
Beyond standalone LMs, we also include two representative repository-level completion pipelines, \textbf{RepoCoder} and \textbf{DraCo}, as inference-time context construction baselines. In our setting, both methods use the same underlying base model \textbf{Qwen2.5-Coder-0.5B}. To ensure a controlled comparison, in caller-available settings we provide the repository as \emph{caller-only} (caller file plus the target function header), whereas in non-caller settings we provide an \emph{empty-repository} variant (target function header only), so that any gain reflects the ability to automatically recover and utilize invocation evidence from the repository rather than additional training.}

This comprehensive set of baselines not only enables us to assess the benefits of introducing calling contexts in training under realistic and competitive settings, but also how our method performs relative to mainstream models under comparable or stronger capacities.

\subsection{Training and Evaluation details.}
\textbf{Training details.}
All training experiments are conducted on a Linux-based workstation equipped with 4 NVIDIA RTX 3090 GPUs (24GB memory each) and 251GB of system RAM. 
We train two families of models: CodeT5 and Qwen2.5-Coder to examine the effect of invocation-aware supervision across different model scales. For CodeT5 models, we adopt the 60M-parameter(small) and 220M-parameter (base) architectures, trained with learning rates of 3e-4 and 2e-4, respectively, using PyTorch with Accelerate and DeepSpeed to support efficient distributed training and enable BF16 mixed precision for improved speed and memory efficiency. For Qwen2.5-Coder-0.5B, we adopt the LLaMA Factory framework~\cite{zheng2024llamafactory}, which provides modular support for decoder-only models and flexible integration with various optimization strategies. We perform continued pretraining on the same invocation-aware dataset, using the Adam optimizer and a cosine decay learning rate schedule with a peak value of 1e-4.

We construct two parallel versions of each model architecture. The invocation-aware version uses training instances that include the target function header, its real callers, and the associated docstring. The context-free version uses the same data sources and training configurations but excludes the caller field from each instance. All other settings remain identical across the two versions, ensuring that the observed differences arise from the inclusion of explicit calling context.

\textbf{Evaluation details.}
All models are evaluated under a unified decoding configuration to ensure reproducibility.
We adopt nucleus sampling (top-$p=0.95$) with temperature $=0.6$.
Random seeds are fixed across all runs to reduce sampling variance.
Unless otherwise specified, each test instance is provided with exactly one caller as its calling context; the $n$-caller analysis in Section~\ref{result:rq4} varies this field by aggregating $n \in \{1, 2,3,\text{all}\}$ callers.
These decoding parameters follow common practice in code generation~\cite{chen2024survey,yu2024codereval}, balancing diversity and determinism.

For our trained models, both training and evaluation use a structured template with optional fields:
the target function header after \texttt{<func>}, the caller context after \texttt{<calledby>}, and a natural-language docstring after \texttt{<docstring>}.
{
For baseline models, we use a \emph{field-wise equivalent} natural-language prompt that exposes the same information when available: it always includes the target header, and includes the caller or docstring only when the corresponding field is enabled by the evaluation setting. Within a fixed setting, different models receive identical inputs. By toggling these two fields across settings, we instantiate distinct prompt configurations with different contextual richness, while keeping the overall serialization format fixed across models.}

\begin{promptbox}{Prompt Template for Caller-Context Code Generation}

\small

\pfield{Task:}
\textit{You are an expert Python programmer. You will be given the header of a function, the body of a function that calls it, and a natural language description explaining what the code should do. Your task is to implement the function so that it integrates correctly with the caller and matches the description. The implementation should be efficient, robust, and handle edge cases. Do not include explanations in your response.}

\pfield{Caller Context (optional)} 
\begin{promptcode}
\pholder{caller}  
\end{promptcode}

\noindent\ldots

\pfield{Target Function (always):}
\begin{promptcode}
\pholder{header}
\end{promptcode}

\pfield{Docstring (optional):}
\begin{promptcode}
\pholder{docstring}  %
\end{promptcode}

\end{promptbox}

\section{Evaluation results}
\label{sec:evaluation}

\begin{table}[t]
\small
\setlength{\tabcolsep}{3pt}
\renewcommand{\arraystretch}{1.2}
\centering
\caption{Performance under caller-available settings.}
\vspace{-1em}
\label{tab:rq1_result}
\begin{adjustbox}{max width=1.\linewidth}
\begin{tabular}{clcccc|cccc}
\toprule
\multicolumn{2}{c}{\multirow{2}{*}{\textbf{Model}}} 
& \multicolumn{4}{c|}{\textsc{CoderEval}} 
& \multicolumn{4}{c}{\textsc{CallerEval}} \\
\cmidrule(lr){3-6} \cmidrule(lr){7-10}
&  & Pass@1 & Pass@5 & CodeBLEU & ROUGE-L & Pass@1 & Pass@5 & CodeBLEU & ROUGE-L\\
\midrule
\multirow{12}{*}{\rotatebox{90}{\textit{Header + Calling}}} 
& CodeGen-Multi-350M & 5.91\% & 16.96\% & 28.21 & 32.15 & 6.92\% & 7.60\% &33.05&37.94\\
& \cellcolor{gray!20}{CodeGen-Multi-2B} & \cellcolor{gray!20}{14.35\%} & \cellcolor{gray!20}{26.09\%} & \cellcolor{gray!20}{29.48} & \cellcolor{gray!20}{32.16} & \cellcolor{gray!20}{10.65\%} & \cellcolor{gray!20}{15.21\%}  & \cellcolor{gray!20}{35.73} & \cellcolor{gray!20}{36.01}\\
& PanGu-300M & 13.91\% & 18.26\% & 24.82 & 31.51 & 8.57\% & 8.93\% & 34.72 & 43.81 \\
& \cellcolor{gray!20}{DeepSeek-Coder-1.3B-base} & \cellcolor{gray!20}{14.52\%} & \cellcolor{gray!20}{31.74\%} & \cellcolor{gray!20}{25.07}  & \cellcolor{gray!20}{23.01} & \cellcolor{gray!20}{7.98\%} & \cellcolor{gray!20}{11.41\%} & \cellcolor{gray!20}{30.16} & \cellcolor{gray!20}{33.71} \\
& \cellcolor{gray!20}{DeepSeek-Coder-6.7B-base} & \cellcolor{gray!20}{31.04}\% & \cellcolor{gray!20}{43.48\%} & \cellcolor{gray!20}{31.32} & \cellcolor{gray!20}{40.25} & \cellcolor{gray!20}{10.72\%} & \cellcolor{gray!20}{15.59\%} & \cellcolor{gray!20}{34.17} & \cellcolor{gray!20}{47.45} \\
& Qwen2.5-Coder-0.5B-Instruct & 12.96\% & 24.35\% &  28.02 & 28.47& 3.27\% & 5.32\% & 29.61  & 33.88 \\
& \cellcolor{gray!20}{Qwen2.5-Coder-1.5B-Instruct} & \cellcolor{gray!20}{28.78\%} & \cellcolor{gray!20}{44.35\%} & \cellcolor{gray!20}{30.23} & \cellcolor{gray!20}{38.59} & \cellcolor{gray!20}{9.51\%} & \cellcolor{gray!20}{12.55\%} &\cellcolor{gray!20}{30.72}& \cellcolor{gray!20}{43.48} \\
& \cellcolor{gray!20}{Qwen2.5-Coder-3B-Instruct}   & \cellcolor{gray!20}{29.39\%} & \cellcolor{gray!20}{44.78\%} &  \cellcolor{gray!20}{31.12} &\cellcolor{gray!20}{39.48}& \cellcolor{gray!20}{12.24\%} & \cellcolor{gray!20}{18.25\%} & \cellcolor{gray!20}{32.26} &  \cellcolor{gray!20}{44.66} \\
& \cellcolor{gray!20}{Qwen2.5-Coder-14B-Instruct}  & \cellcolor{gray!20}{34.78\%} & \cellcolor{gray!20}{49.13\%} &   \cellcolor{gray!20}{32.87} & \cellcolor{gray!20}{41.60} & \cellcolor{gray!20}{13.99\%} & \cellcolor{gray!20}{18.63\%} & \cellcolor{gray!20}{34.14} &  \cellcolor{gray!20}{48.92}\\
& \cellcolor{gray!20}{Qwen2.5-Coder-32B-Instruct}  & \cellcolor{gray!20}{34.26\%} & \cellcolor{gray!20}{48.70\%} & \cellcolor{gray!20}{32.36} &\cellcolor{gray!20}{41.31} & \cellcolor{gray!20}{14.30\%} & \cellcolor{gray!20}{19.67\%} & \cellcolor{gray!20}{33.87} &  \cellcolor{gray!20}{47.41}\\
& \cellcolor{gray!20}{GPT-5.2-Codex}          & \cellcolor{gray!20}{36.96\%} & \cellcolor{gray!20}{52.17\%} & \cellcolor{gray!20}{33.24} & \cellcolor{gray!20}{42.15}& \cellcolor{gray!20}{21.33\%}& \cellcolor{gray!20}{26.92\%} & \cellcolor{gray!20}{38.73} & \cellcolor{gray!20}{51.02} \\

& RepoCoder & 11.66\% & 21.92\% &   25.22& 25.62& 3.79\% & 4.87\% &  30.65 & 33.49 \\
& DraCo & 10.73\% & 20.16\% &  24.16 & 25.11& 3.59\% & 4.40\% & 29.86  & 34.86 \\

\cmidrule{2-10}
& \modelname-small(60M) & 13.91\% & 20.00\%   & 20.19 & 27.84 & 7.81\% & 11.33\%   & 32.72 & 43.81 \\
& \modelname-base(220M) & 15.91\% & 21.30\%   & 21.85 & 31.18 & 13.31\% & 17.49\%  & 35.98 & 45.33 \\
& \textbf{\modelname-qwen(0.5B)} & \textbf{17.74\%} & \textbf{27.39\%}   &  24.99 & 34.01 & \textbf{16.88\%} & \textbf{19.77\%}   & 37.50 & 46.85 \\

\midrule

\multirow{12}{*}{\rotatebox{90}{\textit{Header + Calling + NL}}} & CodeGen-Multi-350M       & 8.78\% & 17.83\%   & 29.75 & 33.43 & 8.97\% & 12.93\%   & 33.92 & 36.28 \\
& \cellcolor{gray!20}{CodeGen-Multi-2B}         & \cellcolor{gray!20}{18.09\%} & \cellcolor{gray!20}{31.30\%}   & \cellcolor{gray!20}{31.30} & \cellcolor{gray!20}{34.12}& \cellcolor{gray!20}{13.54\%} & \cellcolor{gray!20}{18.25\%}   &  \cellcolor{gray!20}{37.59} & \cellcolor{gray!20}{37.25}\\
& PanGu-300M         & 15.22\% & 20.43\%   &  25.35 & 33.38 & 11.79\% & 14.46\% &  38.00 &  49.02 \\
& \cellcolor{gray!20}{DeepSeek-Coder-1.3B-base}        & \cellcolor{gray!20}{17.22\%}      & \cellcolor{gray!20}{32.61\%}   & \cellcolor{gray!20}{24.09} & \cellcolor{gray!20}{20.01} & \cellcolor{gray!20}{12.07\%}& \cellcolor{gray!20}{16.73\%}   & \cellcolor{gray!20}{34.38} & \cellcolor{gray!20}{43.12}\\
& \cellcolor{gray!20}{DeepSeek-Coder-6.7B-base}        & \cellcolor{gray!20}{36.09\%}     & \cellcolor{gray!20}{48.70\%} & \cellcolor{gray!20}{33.17} & \cellcolor{gray!20}{42.52} & \cellcolor{gray!20}{19.01\%}& \cellcolor{gray!20}{23.95\%} & \cellcolor{gray!20}{37.42} & \cellcolor{gray!20}{50.72}\\
& Qwen2.5-Coder-0.5B-Instruct         & 16.26\% & 28.26\% & 30.29&33.31& 6.16\% & 9.51\% & 30.50 & 35.62 \\
& \cellcolor{gray!20}{Qwen2.5-Coder-1.5B-Instruct}         & \cellcolor{gray!20}{32.61\%} & \cellcolor{gray!20}{47.83\%} & \cellcolor{gray!20}{31.67} &\cellcolor{gray!20}{41.40}& \cellcolor{gray!20}{14.07\%}& \cellcolor{gray!20}{16.73\%} & \cellcolor{gray!20}{33.90} & \cellcolor{gray!20}{47.81}\\
& \cellcolor{gray!20}{Qwen2.5-Coder-3B-Instruct}           & \cellcolor{gray!20}{34.09\%} & \cellcolor{gray!20}{48.70\%} & \cellcolor{gray!20}{33.34} &\cellcolor{gray!20}{41.10}& \cellcolor{gray!20}{15.82\%}& \cellcolor{gray!20}{19.39\%} & \cellcolor{gray!20}{35.86} & \cellcolor{gray!20}{48.28}\\
& \cellcolor{gray!20}{Qwen2.5-Coder-14B-Instruct}          & \cellcolor{gray!20}{38.87\%} & \cellcolor{gray!20}{53.04\%} &  \cellcolor{gray!20}{33.78} & \cellcolor{gray!20}{42.53} & \cellcolor{gray!20}{18.71\%}& \cellcolor{gray!20}{22.05\%} & \cellcolor{gray!20}{37.88} & \cellcolor{gray!20}{51.02} \\
& \cellcolor{gray!20}{Qwen2.5-Coder-32B-Instruct}          & \cellcolor{gray!20}{38.17\%} & \cellcolor{gray!20}{50.87\%} & \cellcolor{gray!20}{33.89} & \cellcolor{gray!20}{43.10}& \cellcolor{gray!20}{20.99\%}& \cellcolor{gray!20}{23.57\%} & \cellcolor{gray!20}{38.59} & \cellcolor{gray!20}{53.00} \\
& \cellcolor{gray!20}{GPT-5.2-Codex}          & \cellcolor{gray!20}{42.43\%} & \cellcolor{gray!20}{54.78\%} & \cellcolor{gray!20}{34.73} & \cellcolor{gray!20}{43.88}& \cellcolor{gray!20}{32.55\%}& \cellcolor{gray!20}{37.01\%} & \cellcolor{gray!20}{41.07} & \cellcolor{gray!20}{51.88} \\

& RepoCoder & 16.04\% & 26.70\% & 29.60  & 33.34  & 8.54\% & 11.61 \% & 29.66  & 32.34 \\
& DraCo & 15.08\% & 24.92\% & 28.63  & 33.49 & 7.99\% & 9.21\% & 29.12  & 34.22 \\

\cmidrule{2-10}

& \modelname-small(60M) & 14.96\% & 21.74\% & 21.12 & 29.78 & 11.41\% & 15.40\% & 34.77 & 44.86\\
& \modelname-base(220M) & 21.22\% & 27.83\% & 23.25 & 33.43 & 16.58\% & 20.15\% & 38.34 & 51.60 \\
& \textbf{\modelname-qwen(0.5B)} & \textbf{23.65\%} & \textbf{36.52\%} & 27.26 & 36.47 & \textbf{22.81\%} & \textbf{26.62\%} & 41.28 & 51.80\\

\bottomrule
\end{tabular}
\end{adjustbox}
\end{table}

\subsection{RQ1: Results for Caller-Available Setting}
To assess whether invocation-aware training improves function generation in realistic development scenarios, we begin by evaluating \modelname under caller-available scenario. This aligns with the structure of our training data, where the model is conditioned on actual calling contexts to learn usage-consistent behaviors. In some cases, natural language descriptions are not always present in real code, and when they appear, they often reflect the developer’s thought process during implementation rather than a precise functional specification. We evaluate both settings: with and without such descriptions to reflect this variability and test whether the model can make effective use of natural language. Note that these settings are both possible in real-world development.

As shown in Table~\ref{tab:rq1_result}, for fair comparison,  we shade larger models in gray. First, we focus on models with similar or slightly larger parameter sizes (60M–500M). In the ``Header + Calling" input setting, \modelname-base (220M) achieves a pass@1 of 15.91\% and 13.31\% on \textsc{CoderEval} and \textsc{CallerEval}, respectively. This clearly outperforms CodeGen-350M (5.91\%, 6.92\%) and PanGu-300M (13.91\%, 8.57\%) on the two benchmarks. Our \modelname-small (60M) also matches or outperforms several larger baselines in both settings. Compared with 0.5B–1.3B-scale models such as Qwen2.5-Coder-0.5B-Instruct (12.96\%, 3.27\%) and DeepSeek-Coder-1.3B-base (14.52\%, 7.98\%), \modelname-base consistently remains competitive, highlighting the effectiveness of invocation-aware training in both synthetic and real-world scenarios. {
In addition, we compare against repository-level methods that augment inference-time inputs without changing model weights. \textbf{RepoCoder} (built on \textbf{Qwen2.5-Coder-0.5B-base}) achieves 11.66\% and 3.79\% pass@1 on \textsc{CoderEval} and \textsc{CallerEval}, while \textbf{DraCo} (using the same base model but retrieving definition-centric dependencies) reaches similar pass@1. Compared to directly providing the caller context as input, these retrieval baselines recover part of the invocation evidence but remain below \modelname, indicating that training-time invocation-aware learning yields stronger and more reliable utilization of calling constraints than inference-time context reconstruction alone. Finally, \textbf{GPT-5.2-Codex} attains 36.96\% and 21.33\% pass@1 under the same prompts and metrics, serving as a strong proprietary reference point. CodeBLEU and ROUGE-L provide a complementary reference-based view, their trends are broadly consistent with Pass@K across settings, although they may vary due to the diversity of semantically equivalent implementations.}

When natural language descriptions are included (``Header + Calling + NL"), all models benefit, but \modelname achieves higher or comparable accuracy relative to models with much larger parameters on both benchmarks. For instance, \modelname-base (220M) attains pass@1 scores of 21.22\% on \textsc{CoderEval} and 16.58\% on \textsc{CallerEval}, outperforming CodeGen-2B (18.09\%, 13.54\%) as well as DeepSeek-Coder-1.3B-base (17.22\%, 12.07\%). The largest version, \modelname-qwen (0.5B), achieves 23.65\% and 22.81\% on \textsc{CoderEval} and \textsc{CallerEval} respectively, significantly outperforming Qwen2.5-Coder-0.5B-Instruct (16.26\%, 6.16\%), and narrowing the gap with larger baselines such as Qwen2.5-Coder-3B-Instruct and DeepSeek-Coder-6.7B-base. These results demonstrate that invocation-aware training delivers substantial gains across both benchmarks, especially for small and medium-scale models.

Comparing the results across both datasets, we observe that all models, including \modelname, achieve higher pass@1 scores on \textsc{CoderEval} than on \textsc{CallerEval} under the same settings, reflecting the increased difficulty of context-dependent generation in \textsc{CallerEval}. However, the performance drop for our models, especially \modelname-qwen (0.5B), is notably smaller than that observed for other baselines. For example, \modelname-qwen (0.5B) achieves the best performance among all baseline models on our {\benchmarkname}, reaching 22.81\% pass@1 with full inputs, while its absolute drop from \textsc{CoderEval} (23.65\%) is much smaller than that of Qwen2.5-Coder-0.5B-Instruct (from 16.26\% to 6.16\%). This demonstrates that our invocation-aware training not only improves absolute performance, but also enables the model to maintain robust results as task complexity increases, highlighting its generalization to realistic, context-rich code generation scenarios.

\begin{figure}
    \centering
    \includegraphics[width=0.9\linewidth]{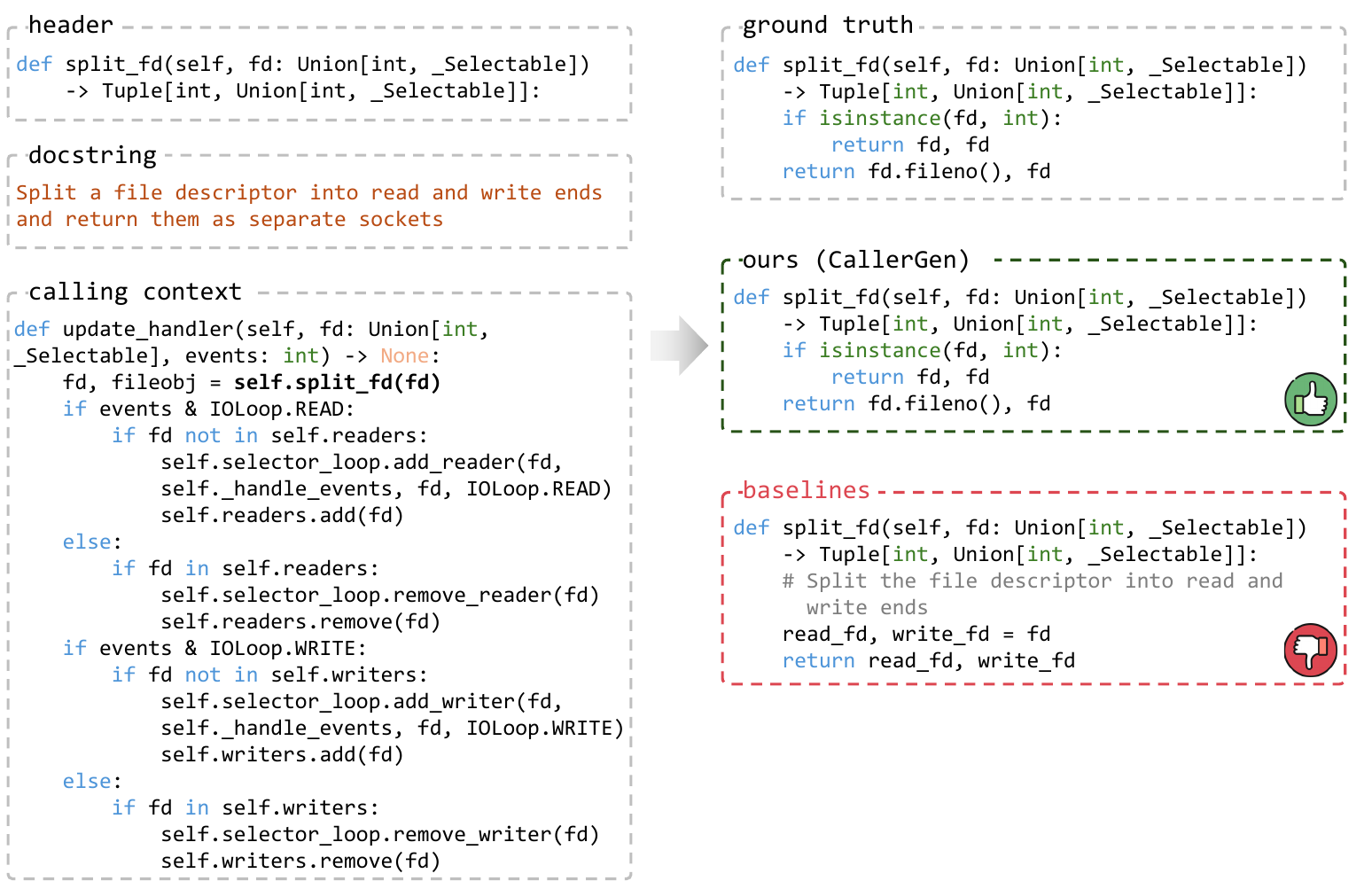}
    \caption{Case 1: a function handles file-descriptor normalization
in an event-driven I/O framework.}
    \label{fig:tornado-case}
\end{figure}

Fig.~\ref{fig:tornado-case} contrasts the baseline and our model on a function that handles file-descriptor normalization in an event-driven I/O framework.  
The natural-language description expresses the intent to ``split a file descriptor into read and write ends,'' which conveys the general purpose but leaves the implementation details underspecified. The caller snippet \texttt{fd, fileobj = self.split\_fd(fd); self.selector\_loop.add\_reader(fd, ...)} clearly treats the first return value fd as a hashable handle reused by \texttt{add\_reader/add\_writer}, and the second as the original object passed along.
The type annotation \texttt{Tuple[int, \_]} and caller-side variable roles precisely define the hidden contract: return (int fd, original object). Our model aligned with this executable contract; the baseline failed to connect the metaphor of the caller’s variable roles with the required return structure. This case exemplifies the abstraction gap between high-level textual intent and concrete behavioral requirements, an ambiguity that caller-side evidence effectively resolves. It further shows that merely providing contextual tokens is insufficient: understanding caller-available dependencies requires model-level ability rather than additional input alone.

\begin{figure}
    \centering
    \includegraphics[width=0.9\linewidth]{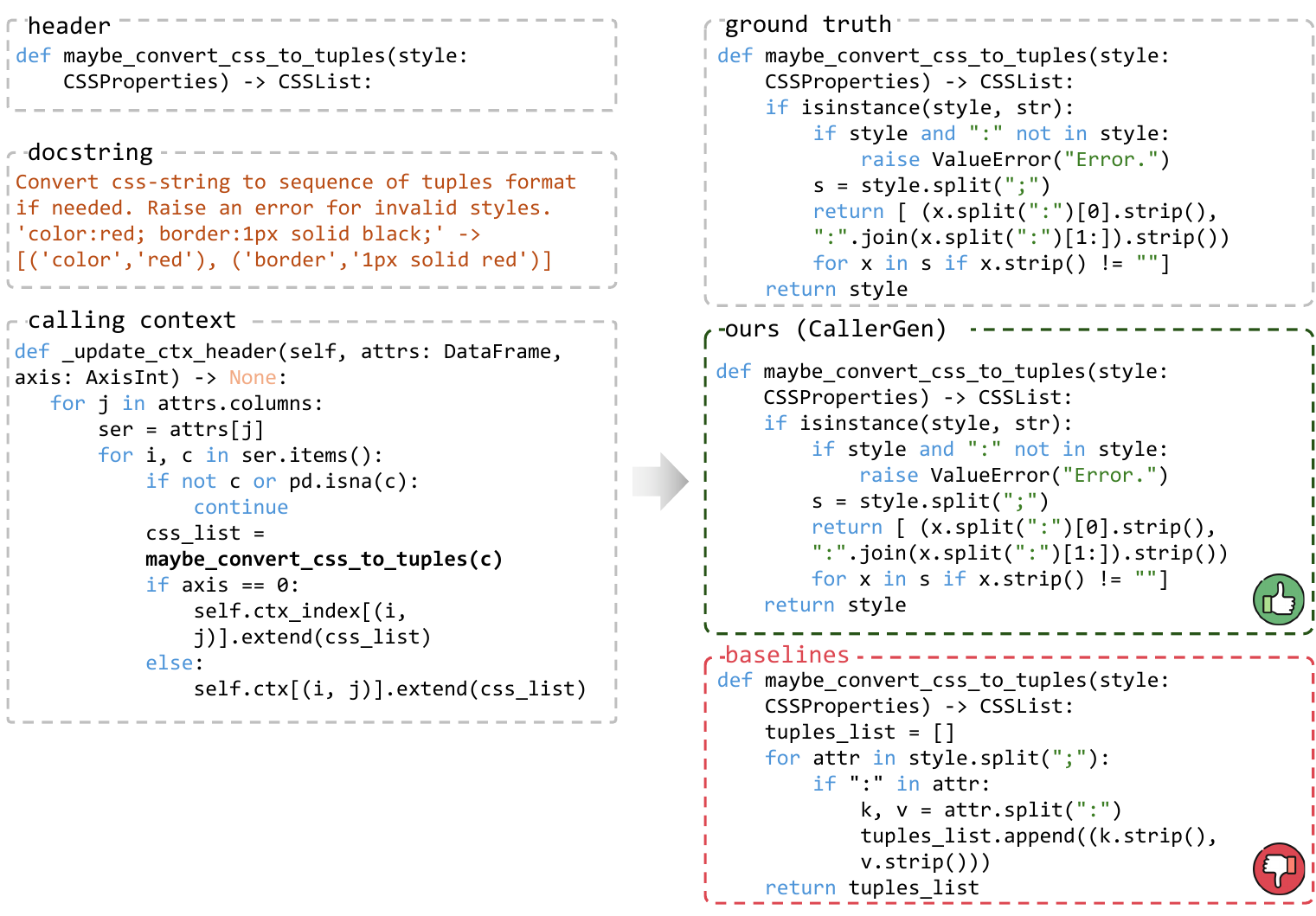}
    \caption{Case 2: a function that converts CSS-style strings into sequences of attribute–value tuples.}
    \label{fig:pandas-case}
\end{figure}

Fig.~\ref{fig:pandas-case} presents a function that converts CSS-style strings into sequences of attribute--value tuples.  The natural-language description correctly outlines the goal but omits structural requirements such as preserving empty attributes and multi-colon values.The caller context, which extends the parsed result into a cumulative style list (\texttt{ctx.extend(css\_list)}), implies that the output must be a flat list of key--value pairs suitable for direct concatenation. Our model correctly inferred this implicit structure and implemented a parsing logic that aggregates all content after the first colon, 
whereas the baseline, though receiving the same context, applied a generic pattern that truncated the values and violated the caller’s expected format.
This contrast demonstrates that effective reasoning over caller evidence depends on the model’s invocation-aware representation, not merely on the presence of surrounding code. Caller context thus serves as a meaningful signal only when the model can internalize and utilize it to satisfy behavioral contracts.

\begin{tcolorbox}[colback=gray!10, colframe=gray!10, sharp corners=southwest, boxrule=0pt, enhanced, borderline west={1pt}{0pt}{gray!60}]
\textbf{Answer to RQ1:} 
\modelname outperforms all baseline models of similar or larger scale on both benchmarks. For example, \modelname-qwen (0.5B) achieves 23.65\% pass@1 on \textsc{CoderEval} and 22.81\% on \textsc{CallerEval}, while the best baseline of the same size, Qwen2.5-Coder-0.5B-Instruct, drops from 16.26\% to 6.16\%. Moreover, the performance drop of \modelname between datasets is much smaller, showing stronger robustness and generalization in realistic, context-rich scenarios.
\end{tcolorbox}

\subsection{RQ2: Results for Non-Caller Setting}
While most functions are written with access to the calling context, there also exist scenarios where functions are implemented directly without such information. One of the most common usage scenarios for coding is to directly implement an unfinished function when only its header is available; another scenario is writing a function based on a natural language description. These non-caller settings are widely represented in existing benchmarks and form a useful basis for evaluating model behavior under limited input conditions.

As shown in Table~\ref{tab:rq2_result}, across non-caller inputs, \modelname demonstrates performance comparable to or exceeding similarly sized models. On header-only inputs, the 60M version already outperforms CodeGen-350M on \textsc{CoderEval}, and the 220M version achieves 10.00\% pass@1, approaching PanGu-300M (11.74\%). On {\benchmarkname}, the gap becomes more apparent: \modelname-220M reaches 7.98\% pass@1, outperforming both CodeGen-350M and PanGu-300M. When evaluated under the commonly adopted header with natural language configuration in prior code generation benchmarks, \modelname maintains strong performance. The 220M model achieves 14.78\% pass@1 on \textsc{CoderEval}, close to CodeGen-350M (16.96\%). Notably, our \modelname-0.5B model trained with invocation-aware supervision outperforms the instruction-tuned Qwen2.5-Coder-0.5B-Instruct on both benchmarks. Through invocation-aware training, \modelname captures general code generation capabilities that enable it to consistently match or outperform similarly sized models across both minimal and NL-included non-caller settings.

\begin{table}[t]
\small
\setlength{\tabcolsep}{3pt}
\renewcommand{\arraystretch}{1.2}
\centering
\caption{Performance under Non-Caller settings.}
\vspace{-1em}
\label{tab:rq2_result}
\begin{adjustbox}{max width=1.\linewidth}
\begin{tabular}{clcccc|cccc}
\toprule
\multicolumn{2}{c}{\multirow{2}{*}{\textbf{Model}}} 
& \multicolumn{4}{c|}{\textsc{CoderEval}} 
& \multicolumn{4}{c}{\textsc{CallerEval}} \\
\cmidrule(lr){3-6} \cmidrule(lr){7-10}
&  & Pass@1 & Pass@5 & CodeBLEU & ROUGE-L & Pass@1 & Pass@5 & CodeBLEU & ROUGE-L \\
\midrule
\multirow{12}{*}{\rotatebox{90}{\textit{Header}}} 
& CodeGen-Multi-350M & 7.39\% & 7.39\% & 21.51 & 24.97 & 3.42\% &  3.68\% & 31.09 &31.94\\
& \cellcolor{gray!20}{CodeGen-Multi-2B}   & \cellcolor{gray!20}{8.70\%} & \cellcolor{gray!20}{8.70\%} & \cellcolor{gray!20}{23.82}&\cellcolor{gray!20}{25.16}& \cellcolor{gray!20}{4.64\%} & \cellcolor{gray!20}{7.22\%} & \cellcolor{gray!20}{29.85} & \cellcolor{gray!20}{33.16} \\
& PanGu-300M   & 11.74\% & 11.74\% & 20.65&27.57 & 4.94\% & 4.94\% & 28.68 & 31.51\\
& \cellcolor{gray!20}{DeepSeek-Coder-1.3B-base} & \cellcolor{gray!20}{8.26\%} & \cellcolor{gray!20}{8.97\%} & \cellcolor{gray!20}{23.89}& \cellcolor{gray!20}{4.20}& \cellcolor{gray!20}{6.69\%} & \cellcolor{gray!20}{6.84\%} & \cellcolor{gray!20}{28.60} & \cellcolor{gray!20}{36.87}\\
& \cellcolor{gray!20}{DeepSeek-Coder-6.7B-base} & \cellcolor{gray!20}{17.30\%} & \cellcolor{gray!20}{17.30\%} &  \cellcolor{gray!20}{20.40} &  \cellcolor{gray!20}{20.25} & \cellcolor{gray!20}{6.84\%} & \cellcolor{gray!20}{6.84\%} & \cellcolor{gray!20}{30.21} & \cellcolor{gray!20}{43.40}\\
& Qwen2.5-Coder-0.5B-Instruct & 10.43\% & 12.17\% & 20.19 &27.23 & 4.03\% & 4.18\% &26.93 & 27.34\\
& \cellcolor{gray!20}{Qwen2.5-Coder-1.5B-Instruct} & \cellcolor{gray!20}{13.48\%} & \cellcolor{gray!20}{13.48\%} &  \cellcolor{gray!20}{20.28} & \cellcolor{gray!20}{27.61} &\cellcolor{gray!20}{5.70\%} & \cellcolor{gray!20}{5.70\%} &\cellcolor{gray!20}{28.35}&\cellcolor{gray!20}{39.06}\\
& \cellcolor{gray!20}{Qwen2.5-Coder-3B-Instruct} & \cellcolor{gray!20}{14.52\%} & \cellcolor{gray!20}{14.76\%} & \cellcolor{gray!20}{20.69} & \cellcolor{gray!20}{28.27} & \cellcolor{gray!20}{9.35\%} & \cellcolor{gray!20}{9.89\%} &\cellcolor{gray!20}{28.98} &\cellcolor{gray!20}{39.06}\\
& \cellcolor{gray!20}{Qwen2.5-Coder-14B-Instruct} & \cellcolor{gray!20}{19.13\%} & \cellcolor{gray!20}{22.61\%} & \cellcolor{gray!20}{20.96}& \cellcolor{gray!20}{29.48}& \cellcolor{gray!20}{9.89\%} & \cellcolor{gray!20}{11.98\%} &\cellcolor{gray!20}{32.05} &\cellcolor{gray!20}{44.66} \\
& \cellcolor{gray!20}{Qwen2.5-Coder-32B-Instruct} & \cellcolor{gray!20}{22.70\%} & \cellcolor{gray!20}{23.04\%} &\cellcolor{gray!20}{20.53} & \cellcolor{gray!20}{29.28} & \cellcolor{gray!20}{9.51\%} & \cellcolor{gray!20}{10.98\%} &\cellcolor{gray!20}{31.52} &\cellcolor{gray!20}{45.98} \\
& \cellcolor{gray!20}{GPT-5.2-Codex}          & \cellcolor{gray!20}{23.30\%} & \cellcolor{gray!20}{24.78\%} & \cellcolor{gray!20}{23.45} & \cellcolor{gray!20}{30.56}& \cellcolor{gray!20}{19.05\%}& \cellcolor{gray!20}{21.83\%} & \cellcolor{gray!20}{34.07} & \cellcolor{gray!20}{45.52} \\
& RepoCoder & 9.39\% & 10.95\% & 18.17  & 24.51 & 3.63\% & 3.76\% & 24.24  & 24.61 \\
& DraCo & 9.47\% & 10.88\% &  18.09 & 24.63 & 3.58\% & 3.79\% & 24.12  & 24.73 \\

\cmidrule{2-10}
& \modelname-small(60M) & 9.56\% & 9.56\% & 18.42 & 24.52 & 4.41\% & 4.41\% & 30.51& 38.80\\
& \modelname-base(220M) & 10.00\% & 10.00\% & 19.62 &27.14& 7.98\% & 9.02\% & 33.87 & 44.96\\
& \textbf{\modelname-qwen(0.5B)} & \textbf{12.61\%} & \textbf{13.48\%} & \textbf{18.38} & \textbf{26.01} & \textbf{9.51\%} & \textbf{10.84\%} & 35.42 & 45.10\\

\midrule

\multirow{12}{*}{\rotatebox{90}{\textit{Header + NL}}}
& CodeGen-Multi-350M       & 16.96\% & 18.39\% & 23.87 & 27.90 & 9.13\% & 11.79\% & 32.79& 34.70\\
& \cellcolor{gray!20}{CodeGen-Multi-2B}         & \cellcolor{gray!20}{20.87\%} & \cellcolor{gray!20}{28.57\%}& \cellcolor{gray!20}{25.61}&\cellcolor{gray!20}{28.14}& \cellcolor{gray!20}{11.41\%} & \cellcolor{gray!20}{15.02\%} & \cellcolor{gray!20}{36.23} & \cellcolor{gray!20}{32.06}\\
& PanGu-300M         & 17.39\% & 19.05\% & 22.69&31.81 & 10.71\% & 12.13\% & 30.86 & 34.65 \\
& \cellcolor{gray!20}{DeepSeek-Coder-1.3B-Base}        & \cellcolor{gray!20}{10.43\%} & \cellcolor{gray!20}{12.17\%} & \cellcolor{gray!20}{21.95} & \cellcolor{gray!20}{16.04} & \cellcolor{gray!20}{13.08\%}& \cellcolor{gray!20}{13.31\%}&\cellcolor{gray!20}{34.23} &\cellcolor{gray!20}{44.95}\\
& \cellcolor{gray!20}{DeepSeek-Coder-6.7B-Base}        & \cellcolor{gray!20}{25.30\%} & \cellcolor{gray!20}{25.65\%} & \cellcolor{gray!20}{25.31} & \cellcolor{gray!20}{33.98} & \cellcolor{gray!20}{16.88\%}& \cellcolor{gray!20}{17.11\%}&\cellcolor{gray!20}{35.74}&\cellcolor{gray!20}{48.47}\\
& Qwen2.5-Coder-0.5B-Instruct         & 14.78\% & 17.83\% & 25.72 & 33.54 & 9.13\% & 9.13\%& 29.45 & 31.52\\
& \cellcolor{gray!20}{Qwen2.5-Coder-1.5B-Instruct}         & \cellcolor{gray!20}{25.65\%} & \cellcolor{gray!20}{28.70\%} & \cellcolor{gray!20}{23.14} & \cellcolor{gray!20}{32.49} & \cellcolor{gray!20}{12.70\%}& \cellcolor{gray!20}{12.93\%}& \cellcolor{gray!20}{33.58} & \cellcolor{gray!20}{44.49}\\
& \cellcolor{gray!20}{Qwen2.5-Coder-3B-Instruct}           & \cellcolor{gray!20}{28.87\%} &  \cellcolor{gray!20}{29.57\%}& \cellcolor{gray!20}{24.44} & \cellcolor{gray!20}{32.61} & \cellcolor{gray!20}{13.99\%}& \cellcolor{gray!20}{14.07\%}& \cellcolor{gray!20}{34.95} & \cellcolor{gray!20}{42.84}\\
& \cellcolor{gray!20}{Qwen2.5-Coder-14B-Instruct}          & \cellcolor{gray!20}{31.04\%} & \cellcolor{gray!20}{37.30\%} & \cellcolor{gray!20}{25.17}& \cellcolor{gray!20}{34.52}& \cellcolor{gray!20}{16.35\%}& \cellcolor{gray!20}{16.35\%}&\cellcolor{gray!20}{36.23}& \cellcolor{gray!20}{47.89}\\
& \cellcolor{gray!20}{Qwen2.5-Coder-32B-Instruct}          & \cellcolor{gray!20}{30.35\%} & \cellcolor{gray!20}{37.43\%} & \cellcolor{gray!20}{25.60} & \cellcolor{gray!20}{34.51} & \cellcolor{gray!20}{17.11\%}& \cellcolor{gray!20}{18.21\%}& \cellcolor{gray!20}{36.09} &\cellcolor{gray!20}{50.04} \\
& \cellcolor{gray!20}{GPT-5.2-Codex}          & \cellcolor{gray!20}{35.59\%} & \cellcolor{gray!20}{38.24\%} & \cellcolor{gray!20}{26.43} & \cellcolor{gray!20}{35.23}& \cellcolor{gray!20}{24.55\%}& \cellcolor{gray!20}{30.42\%} & \cellcolor{gray!20}{37.69} & \cellcolor{gray!20}{50.72} \\
& RepoCoder & 13.30\% & 16.05\% &  23.15 & 30.19& 8.22\% & 8.22\% &  26.51 & 28.37 \\
& DraCo & 13.10\% & 16.76\% &  23.75 & 30.29& 8.35\% & 8.35\% &  26.70 & 26.92 \\

\cmidrule{2-10}
& \modelname-small(60M) & 12.17\% & 14.78\% & 20.21 & 27.54 & 6.56\% & 7.66\% & 31.96 & 40.21\\
& \modelname-base(220M) & 14.78\% & 17.83\% & 21.35 & 29.82 & 12.93\% & 13.69\% & 36.21 & 48.72 \\
& \textbf{\modelname-qwen(0.5B)} & 16.83\% & 18.09\% & 20.28 & 27.75 & \textbf{13.92\%} & \textbf{14.07\%} & 37.58 & 49.53\\

\bottomrule
\end{tabular}
\end{adjustbox}
\end{table}

Compared to larger-scale models, \modelname-0.5B achieves competitive performance under non-caller settings, often narrowing the gap with models several times its size. On {\benchmarkname}, \modelname-0.5B delivers comparable results to Qwen2.5-Coder-3B-Instruct. These results suggest that the benefits of invocation-aware training extend beyond its caller-available scenario. \modelname not only leverages calling contexts effectively, but also integrates natural language descriptions to produce behaviorally aligned and functionally correct code. This complementary use of invocation and NL signals demonstrates its robustness. Even when the calling context is absent, our 0.5B model achieves results comparable to substantially larger baselines, underscoring the broad applicability of the learned generation capabilities.

While \modelname achieves competitive or better results in non-caller settings, we observe that all models, including our own, experience a noticeable performance drop compared to the caller-available scenario discussed in RQ1. For example, \modelname-qwen (0.5B) achieves 22.81\% pass@1 on \textsc{CallerEval} with full caller context, but only 13.92\% under header+NL inputs. Similarly, DeepSeek-Coder-6.7B-Base drops from 19.01\% to 16.88\%. This trend holds across both \textsc{CoderEval} and \textsc{CallerEval}, and for all model sizes. These results highlight that the lack of invocation context leads to a substantial decrease in generation quality for all methods. Therefore, calling context is crucial for producing accurate and context-aware code, and should be considered a necessary component in the design of practical code generation.

\begin{tcolorbox}[colback=gray!10, colframe=gray!10, sharp corners=southwest, boxrule=0pt, enhanced, borderline west={1pt}{0pt}{gray!60}]
\textbf{Answer to RQ2:} \modelname achieves good performance in non-caller settings, outperforming or matching similarly sized and even larger models (e.g., 13.92\% pass@1 for \modelname-qwen (0.5B) on \textsc{CallerEval}). However, all models experience a clear performance drop compared to the caller-available setting, indicating that calling context is essential for accurate and context-aware code generation. %

\end{tcolorbox}

\subsection{RQ3: Benefits across Model Architectures}
After demonstrating the overall advantages of our invocation-aware training approach, we further analyze whether these gains persist across different model architectures and training configurations.
For each base model (CodeT5-small-60M, CodeT5-base-220M, and Qwen2.5-coder-0.5B), we compare the performance between two versions: one that incorporates explicit calling context during training and another that omits it while keeping all other factors unchanged.

\begin{table*}[b]
\centering
\caption{Pass@1 performance of \textsc{CallerGen} variants under different input configurations on \textsc{CoderEval} and \textsc{CallerEval}.}
\label{tab:rq4_result}
\small
\setlength{\tabcolsep}{4pt}
\begin{adjustbox}{max width=\textwidth}
\begin{tabular}{lcccccccc}
\toprule
\textbf{Model} & \multicolumn{4}{c}{\textsc{CoderEval}} & \multicolumn{4}{c}{\textsc{CallerEval}} \\
\cmidrule(lr){2-5} \cmidrule(lr){6-9}
& Header & Header + NL & Header + Caller & Header + Caller + NL 
& Header & Header + NL & Header + Caller & Header + Caller + NL \\
\midrule
w/o calling context(60M)     & 7.48\%  & 11.91\% & 9.13\% & 12.17\% & 3.27\%  & 5.73\%  & 5.51\%  & 7.22\%  \\
\modelname-small     & 9.56\% & 12.17\% & 13.91\% & 14.96\% & 4.41\%  & 6.56\%  & 7.81\%  & 11.41\%  \\
\midrule
w/o calling context(220M)     & 7.83\% & 13.04\% & 11.57\% & 15.57\% & 3.88\%  & 7.22\% & 6.58\% & 10.72\% \\
\modelname-base(220M)     & 10.00\% & 14.78\% & 15.91\% & 21.22\% & 7.98\%  & 12.93\% & 13.31\% & 16.58\% \\
\midrule
w/o calling context(0.5B)     & 11.74\% & 15.22\% & 13.48\% & 18.09\% & 5.86\%  & 11.03\% & 10.41\% & 12.13\% \\
\modelname-qwen(0.5B)     & 12.61\% & 16.83\% & 17.74\% & 23.65\% & 9.51\%  & 13.92\% & 16.88\% & 22.81\% \\
\bottomrule
\end{tabular}
\end{adjustbox}
\end{table*}

As shown in Table~\ref{tab:rq4_result}, \modelname consistently outperforms its non-invocation-aware counterparts across all model architectures and input settings on both \textsc{CallerEval} and \textsc{CoderEval}. For instance, with CodeT5-base (220M) on \textsc{CallerEval}, \modelname achieves a pass@1 of 16.58\% with full inputs (Header + Caller + NL), compared to 10.72\% for the omitted calling context training model, improvement of over 6 percentage points. Similar trends are observed for the 0.5B-scale models: \modelname-qwen-coder achieves 22.81\% pass@1, substantially higher than the Qwen2.5-coder training without calling context (12.13\%) under the same configuration. Even in more constrained settings such as header-only or header+NL, invocation-aware training brings consistent gains. Notably, the improvements are especially pronounced when caller context is available. For example, with CodeT5-base (220M) on \textsc{CallerEval}, the gain from 12.93\% (Header + NL) to 16.58\% (Header + Caller + NL) demonstrates that the model benefits more from contextual information when trained in an invocation-aware manner. 
Similar relative gains are seen across all model sizes and on both benchmarks, indicating that our training approach is robust and generalizes well to different architectures.

\begin{tcolorbox}[colback=gray!10, colframe=gray!10, sharp corners=southwest, boxrule=0pt, enhanced, borderline west={1pt}{0pt}{gray!60}]
\textbf{Answer to RQ3:} 
Invocation-aware training consistently improves performance over standard training across all model architectures. For example, on \textsc{CallerEval} with full inputs, \modelname-base (220M) achieves 16.58\% pass@1 compared to 10.72\% for CodeT5-base training without calling context, and \modelname-qwen (0.5B) achieves 22.81\% versus 12.13\%. The improvements are particularly evident with caller context, showing that invocation-aware models make better use of contextual information across backbones.

\end{tcolorbox}

{
\subsection{RQ4: What aspects of caller context matter}
\label{result:rq4}
To analyze what aspects of caller context matter, we note that the provided context is not a single signal.  It can include the invocation statement, how the return value is used, and surrounding control structures that govern execution.  The amount of caller evidence may also affect performance when multiple callers are available.  To capture these effects, we analyze caller context along two dimensions.  A component ablation isolates different semantic cues to identify which parts of the caller context account for the gains. We further examine how performance changes when providing multiple callers at inference time and when augmenting training with more callers and deeper caller chains. All analyses in this section are conducted on both CoderEval and CallerEval using the 0.5B backbone model. In all variants, each input contains the callee function header together with the caller-side context and natural language description. 
}

{
In the component ablation, we replace the full caller-context snippet with a targeted component extracted from the caller to isolate the contribution of each evidence type. Consider four semantic variants that selectively expose different types of caller information: (i) Caller-signature-only, we provide only the caller function header. This variant captures interface-level and naming cues from the caller while removing executable evidence around the invocation; (ii) Call-site-only, we provide only the call statement that invokes the callee, without surrounding statements; (iii) Data-flow, we provide the call statement together with the immediate uses of its return value in the caller, such as assignments, downstream passing, indexing, or conditional checks. These usage patterns impose direct constraints on the expected behavior of the callee; (iv) Control-flow, we provide the call statement together with invocation-relevant control structures. Concretely, we extract structured blocks such as conditional branches, loops, and exception-handling constructs. When no invocation-relevant structured control evidence can be extracted, we fall back to using the original caller snippet for that instance to avoid empty context. Beyond these semantic components, we introduce two control conditions to test whether the gains can be attributed to non-semantic factors, such as prompt length or superficial surface forms: (v) Length-matched irrelevant context, we replace the caller-side content with a format-preserving snippet of similar length that is unrelated to the target invocation. This controls for token budget and formatting. (vi) Semantics-preserving perturbation, we apply safe rewrites that keep the semantic evidence intact while perturbing surface forms, such as systematic renaming and equivalent syntactic rewrites. This tests whether the model relies on incidental lexical cues.
}

\begin{table}[t]
\small
\setlength{\tabcolsep}{3pt}
\renewcommand{\arraystretch}{1.2}
\centering
\caption{Performance under different caller variant settings.}
\vspace{-1em}
\label{tab:rq4_result}
\begin{adjustbox}{max width=1.\linewidth}
\begin{tabular}{lcccc|cccc}
\toprule
\multirow{2}{*}{\textbf{caller variant}}
& \multicolumn{4}{c|}{\textsc{CoderEval}} 
& \multicolumn{4}{c}{\textsc{CallerEval}} \\
\cmidrule(lr){2-5} \cmidrule(lr){6-9}
& Pass@1 & Pass@5 & CodeBLEU & ROUGE-L & Pass@1 & Pass@5 & CodeBLEU & ROUGE-L \\
\midrule
Caller-signature-only
& 15.74\% & 18.35\% & 21.33 & 30.87 
& 13.99\% & 15.59\% & 39.71 & 48.38 \\
Caller-site-only
& 17.04\% & 18.35\% & 22.50 & 31.13 
& 17.90\% & 18.73\% & 40.26 & 48.34 \\
Data-flow
& 18.87\% & 24.43\% & 22.89 & 32.26 
& 19.13\% & 20.35\% & 41.22 & 49.11 \\
Control-flow
& 19.65\% & 29.00\% & 23.24 & 32.47
& 21.56\% & 25.29\% & 41.93 & 50.28 \\
Length-matched irrelevant
& 17.74\% & 21.83\% & 21.71 & 30.70 
& 11.86\% & 14.59\% & 36.85 & 45.14 \\
Semantics preserving
& 22.61\% & 30.52\% & 24.29 & 34.16 
& 21.79\% & 25.29\% & 41.23 & 50.02 \\
\midrule
\textbf{Full caller}
& \textbf{23.65\%} & \textbf{36.52\%} & 27.26 & 36.47 
& \textbf{22.81\%} & \textbf{26.62\%} & 41.28 & 51.80 \\
\bottomrule
\end{tabular}
\end{adjustbox}
\end{table}

{
As shown in Table~\ref{tab:rq4_result}, we observe three consistent findings across both benchmarks. First, interface-level cues from the caller are not sufficient.  The caller-signature-only variant stays close to the non-caller baseline on both benchmarks, indicating that type-like and naming priors from the caller provide at offer limited guidance for implementing the callee. }

{
Second, we find that exposing how the invocation is used in the caller is more beneficial than exposing the invocation itself. Both the data-flow and control-flow variants outperform the call-site-only variant, indicating that the model benefits primarily when the context provides actionable behavioral constraints tied to the invocation. The call-site alone reveals that an invocation exists, but it often under-specifies what the callee must do. In contrast, return-value usage and invocation-relevant control evidence specify how the callee’s output influences downstream computation, which helps the model produce behaviorally aligned implementations. }

{
To contextualize the control-flow signal, we quantify how frequently invocation-related control evidence occurs in the two benchmarks.    On CallerEval, structured control patterns around invocations are common: for example, call-sites enclosed by a structured block account for 30.12\%  instances and appear in 56.27\%  tasks, indicating that control-flow cues are available for a majority of tasks. Meanwhile, cases where the call’s return value later influences a structured block are less frequent, covering 5.41\%  instances and appearing in 15.21\%  tasks, suggesting that this particularly strong form of control dependence is present but not ubiquitous. On CoderEval, structured blocks that enclose the call-site are present in 14.70\%  instances and 25.22\%  tasks, and callers that contain control flow unrelated to the target invocation account for 18.17\%  instances and appear in 31.74\%  tasks.  At the same time,  a large portion of instances fall into the case where no structured control can be identified, accounting for 59.83\%  instances and appearing in 75.22\%  tasks.  These statistics indicate that control-flow evidence is common but varies in how it relates to the target function’s return value.}

{
Third, to address the potential risk that neural models may rely on surface-level regularities correlated with training outcomes~\cite{ye2024clever, veitch2021counterfactual}, we provide additional evidence that the improvements stem from invocation-relevant semantic cues rather than superficial properties of the input. When the caller-side content is replaced with a format-preserving snippet of similar length but unrelated to the target invocation, the performance remains clearly below the variants that expose return-value usage or invocation-relevant control evidence. This suggests that the main benefit is not driven by simply including more code in the prompt. Moreover, when we keep the same semantic evidence while perturbing surface forms through safe rewrites, the performance changes little, indicating that the model does not depend on incidental lexical or formatting details. Together, these results support that caller context helps primarily because it conveys stable behavioral constraints tied to how the invocation is used, which leads to more behaviorally aligned implementations.}

{
We next study how performance changes as we use different forms of caller organization during training and aggregate multiple callers at inference time. We conduct this study using \textsc{Qwen2.5-Coder-0.5B} and train four variants: three multi-caller variants with $n_{\text{train}}\in\{1,2,3\}$, a two-hop variant that augments each one-hop caller with its immediate caller when available. When $n_{\text{train}}>1$, we keep the original real caller for each callee and sample $(n_{\text{train}}-1)$ additional callers from the remaining available callers of the \emph{same} callee.%
At inference time, we independently vary $n_{\text{test}}$ by constructing inputs that aggregate $n_{\text{test}}\in\{1,2,3,\text{all}\}$ callers for each test item. We report the full performance matrix on both \textsc{CoderEval} and \textsc{CallerEval}, which quantifies the marginal benefit of adding callers and reveals when additional caller evidence saturates or becomes noisy.
}

{
As shown in Table~\ref{tab:ncaller_matrix}, we observe two consistent effects. 
Increasing $n_{\text{train}}$ improves the model's \emph{training-time} ability to utilize caller evidence, and this benefit becomes most apparent when richer caller evidence is available at evaluation.
On \textsc{CoderEval}, under the coverage-maximizing $n_{\text{test}}{=}\text{all}$ condition, pass@1 increases monotonically from 25.59\% to 27.57\% and 28.52\% as $n_{\text{train}}$ increases from 1 to 2 and 3.
A complementary view is the gain from expanding inference-time evidence from a single caller to all callers: this gain grows from +1.94 points ($n_{\text{train}}{=}1$: 23.65\% $\rightarrow$ 25.59\%) to +5.74 points ($n_{\text{train}}{=}2$: 21.83\% $\rightarrow$ 27.57\%) and +4.95 points ($n_{\text{train}}{=}3$: 23.57\% $\rightarrow$ 28.52\%).
This widening gap indicates improved evidence utilization learned during training, rather than a purely prompt-level effect.
For a fixed $n_{\text{train}}$, aggregating more callers generally improves performance (e.g., for $n_{\text{train}}{=}1$, 23.65\% $\rightarrow$ 24.20\% $\rightarrow$ 24.74\% $\rightarrow$ 25.59\% as $n_{\text{test}}$ increases from 1 to 2, 3, and all on \textsc{CoderEval}). In contrast, the two-hop variant underperforms across all $n_{\text{test}}$ on both benchmarks, suggesting that our current multi-hop construction introduces noise and consumes context budget without providing reliable additional constraints.}

{
While multi-caller supervision is beneficial overall, its effect is not strictly monotonic for every fixed $n_{\text{test}}$, suggesting diminishing returns and an evidence-budget mismatch effect.
This is expected because additional callers for the same callee often convey overlapping invocation semantics (redundant argument conventions or repeated return-value usage), so increasing $n_{\text{train}}$ may add limited new constraints while requiring the model to integrate longer and more heterogeneous caller snippets. For example, when evaluated with $n_{\text{test}}{=}3$, $n_{\text{train}}{=}2$ achieves the best performance (26.09\%), while $n_{\text{train}}{=}3$ is slightly lower (25.30\%).
Similarly, under the single-caller evaluation regime ($n_{\text{test}}{=}1$), $n_{\text{train}}{=}2$ can underperform $n_{\text{train}}{=}1$ (21.83\% vs.\ 23.65\%).
These patterns suggest that multi-caller training learns an aggregation prior that is most effective when the inference-time evidence budget is aligned (or moderately aligned) with the training-time budget. 
}

{
We include $n_{\text{test}}{=}\text{all}$ as a coverage-maximizing test that approximates the scenario where all available one-hop calling contexts can be provided.
On \textsc{CoderEval}, $n_{\text{test}}{=}\text{all}$ yields the best performance for all $n_{\text{train}}$ values (25.59\%, 27.57\%, and 28.52\%), confirming that richer caller evidence is beneficial when the context budget permits.
Importantly, the improvement from increasing $n_{\text{train}}$ persists under this strongest evidence budget, showing that training-time gains transfer to richer inference-time evidence and are not limited to the default one-caller evaluation.
In practice, concatenating all callers may exceed the context budget or substantially increase inference cost; therefore, we use a budgeted setting in the main experiments (e.g., one caller per instance) for controlled comparison, and report $n_{\text{test}}{=}\text{all}$ as an additional analysis when feasible.}

{
Overall, training is crucial for utilizing caller evidence.
Providing more callers at inference time is not sufficient by itself; rather, the model must learn during training how to interpret and reconcile invocation constraints across callers.
On \textsc{CoderEval}, expanding the inference-time evidence budget from one caller to all callers yields a +1.94 point gain when trained with $n_{\text{train}}{=}1$, whereas the gain increases to +5.74 points for $n_{\text{train}}{=}2$ and +4.95 points for $n_{\text{train}}{=}3$.
This increasing return on additional caller evidence indicates improved utilization efficiency learned from multi-caller supervision.
At the same time, the full $(n_{\text{train}},n_{\text{test}})$ matrix reveals a budget-mismatch effect, where the best $n_{\text{train}}$ is not strictly monotonic for every fixed $n_{\text{test}}$, suggesting that the optimal aggregation strategy depends on the expected caller-availability budget at deployment.
Accordingly, we use the one-caller setting as the default in our main experiments to ensure controlled and reproducible comparisons under a fixed context budget, and report the $n$-caller matrix as a complementary analysis that characterizes how the learned capability scales when additional calling contexts are available. }

\begin{tcolorbox}[colback=gray!10, colframe=gray!10, sharp corners=southwest, boxrule=0pt, enhanced, borderline west={1pt}{0pt}{gray!60}]
\textbf{Answer to RQ4:} 
\noindent\textbf{Answer to RQ4.}
Caller context helps primarily through invocation-level semantic evidence rather than prompt length alone.
Our component ablation shows that providing richer caller semantics substantially improves functional correctness: on \textsc{CallerEval}, \textsc{Full caller} achieves 22.81\% pass@1 (26.62\% pass@5), far above \textsc{Caller-site-only} at 17.90\% pass@1. The same trend holds on \textsc{CoderEval}, where \textsc{Full caller} reaches 23.65\% pass@1 (36.52\% pass@5), compared to 17.04\% pass@1 for \textsc{Caller-site-only} and 15.74\% for \textsc{Caller-signature-only}. Beyond cue types, multi-caller supervision improves training-time utilization of caller evidence. On \textsc{CoderEval}, under the coverage-maximizing $n_{\text{test}}{=}\text{all}$ setting, pass@1 increases monotonically from 25.59\% to 27.57\% and 28.52\% as $n_{\text{train}}$ increases from 1 to 2 and 3. Moreover, the benefit of expanding inference-time evidence from one caller to all callers grows with multi-caller training (e.g., +2.78 points for $n_{\text{train}}{=}1$ vs.\ +5.74 for $n_{\text{train}}{=}2$), indicating that the model becomes more effective at exploiting caller evidence due to training rather than merely receiving longer prompts.
\end{tcolorbox}

\begin{table}[t]
\small
\setlength{\tabcolsep}{3pt}
\renewcommand{\arraystretch}{1.2}
\centering
\caption{pass@1 of aggregating multiple callers during training and inference time.}
\vspace{-1em}
\label{tab:ncaller_matrix}
\begin{tabular}{lcccc|cccc}
\toprule
\multirow{2}{*}{}
& \multicolumn{4}{c|}{\textsc{CoderEval}} 
& \multicolumn{4}{c}{\textsc{CallerEval}} \\
\cmidrule(lr){2-5} \cmidrule(lr){6-9}
& $n_{\text{test}}{=}1$ & $n_{\text{test}}{=}2$ & $n_{\text{test}}{=}3$ & $n_{\text{test}}{=}\text{all}$
& $n_{\text{test}}{=}1$ & $n_{\text{test}}{=}2$ & $n_{\text{test}}{=}3$ & $n_{\text{test}}{=}\text{all}$ \\
\midrule
\textbf{$n_{\text{train}}$}=1 & 23.65\% & 24.20\% & 24.74\% & 25.59\% & \textbf{22.81\%} & 23.42\% & 23.98\% & 23.65\% \\
\textbf{$n_{\text{train}}$}=2 & 21.83\% & 24.00\% & \textbf{26.09\%} & 27.57\% & 21.57\% & \textbf{24.05\%} & 22.15\% & 23.95\% \\
\textbf{$n_{\text{train}}$}=3 & \textbf{23.57\%} & \textbf{25.48\%} & 25.30\% & \textbf{28.52\%} & 21.79\% & 23.91\% & \textbf{26.12\%} & \textbf{24.81\%} \\
\textbf{$multi-hop$} & 21.13\% & 23.81\% & 24.22 & 24.87\% & 20.98\% & 21.39\% & 22.11\% & 21.98\% \\

\bottomrule
\end{tabular}
\end{table}

\section{Discussion} 
{
Calling contexts provide valuable and actionable behavioral signals that directly reflect developer intent. These signals commonly exist in real-world codebases and offer a more reliable and abundant source of training compared to the vague and sparse natural language descriptions. This widespread availability makes invocation-aware training effective for modeling realistic code generation, as it can be applied consistently across a variety of software projects. More importantly, calling contexts capture how a function is actually invoked and what constraints are imposed by its usage, which helps models generate implementations that better match real-world expectations. Since calling contexts can be extracted directly from existing code, this approach remains effective and feasible at scale without requiring extensive manual annotation.}

{
A related concern is whether neural models truly rely on invocation-relevant semantics, as opposed to learning shortcuts that correlate with success in a particular training or evaluation setup. Spurious correlations are widely observed in machine learning systems and can lead models to depend on non-essential cues that do not reflect the intended semantics~\cite{ye2024clever}. Recent work on language models shows that their outputs can be sensitive to surface rephrasings even when meaning is preserved. Other studies further suggest that models may latch onto syntactic templates as domain cues, which can bias generation beyond what the underlying semantics would warrant~\cite{shaib2025learning}. Motivated by this line of work, we therefore include two robustness-oriented variants to assess whether the observed gains depend on invocation-relevant evidence or on superficial properties of the input in Section~\ref{result:rq4}. Specifically, one variant keeps the caller-side context length and formatting comparable while removing invocation relevance, which mirrors counterfactual-style stress testing where irrelevant attributes are perturbed while controlling other factors~\cite{veitch2021counterfactual, bai2025invariant}. Another variant applies semantics-preserving rewrites to caller evidence, aligning with surface-invariant evaluation principles for LLMs~\cite{cohen2025forget} and with metamorphic testing practices that use semantics-preserving transformations to assess robustness of models~\cite{asgari2025metamorphic, cho2025metamorphic}. Across both benchmarks, the observed pattern is consistent with a semantics-driven interpretation: gains diminish when invocation relevance is removed despite comparable context length, and they remain largely stable under semantics-preserving rewrites. Together with the component ablation showing that return-value usage and invocation-relevant control evidence contribute more than interface-level cues, these results suggest that calling contexts help primarily by providing robust behavioral constraints tied to how a function is used in practice.
}

{
A further practical factor is that calling context may be imperfect: callers can be incomplete, buggy, or reflect incidental local conventions that should not be treated as hard requirements. Such noise can bias any context-conditioned generation toward over-specific behavior or even propagate caller-side mistakes.  In our formulation, calling context is therefore treated as usage cues rather than a complete specification, and generation is still guided by other available signals such as the function signature and optional documentation.  We also rely on data-quality controls in corpus construction (e.g., executability and filtering of incomplete, invalid and test-related code) and on execution-based validation during evaluation to reduce the impact of low-quality supervision.  More broadly, improving robustness to noisy callers motivates stronger data-quality checks and post-generation validation as natural directions to strengthen practical reliability.
}

{
At the same time, the gains from invocation-aware training are not uniform across settings. They depend on (i) how the calling context is organized and how much of it is available, (ii) how a benchmark exposes the calling context with different task difficulty, and (iii) the availability and fidelity of extracted invocation relations in a given language and ecosystem. We discuss these factors below to help interpret variations in absolute performance and relative improvements. For the first point, one-hop calling context from a single caller is a minimal and controllable unit of invocation information. Combining multiple callers is a natural way to impose stronger joint constraints than any single call site alone. However, such multi-caller context is not always available in practice due to the repository stage, extraction coverage limits. At the same time, directly concatenating multiple call sites into a single prompt can introduce confounding effects from context-budget allocation, truncation, and ordering, and it may also blend caller-specific incidental details in ways that increase ambiguity. Likewise, richer context modalities such as multi-hop call chains, dependency graphs, and execution traces can expose deeper program semantics, such as transitive dependencies, data or control-flow constraints, and runtime-observed behaviors. These modalities represent a complementary extension axis to one-hop context: they enrich the modality of conditioning signals and can be combined with caller multiplicity when available. However, this requires reliable extraction and evaluation protocols that isolate the contribution of each modality. Invocation context availability also depends on project organization. In many codebases, established framework or library APIs are used as external dependencies, so call sites are observable and informative as usage context, while the corresponding implementations reside outside the current project scope. Meanwhile, projects often include internal library-like modules or public interfaces that are implemented before local call sites appear (or when callers live outside the current code boundary). In such callee-first cases, invocation-conditioned inputs are naturally unavailable and generation relies on the remaining signals (e.g., signature and optional documentation), corresponding to the non-caller setting reported in Table~4.}

{
These differences further interact with benchmark design and task construction. Both \textsc{CoderEval} and \benchmarkname are evaluated in Python, yet they differ in how the invocation context is specified and how strongly tasks are connected to observable call sites. \textsc{CoderEval} does not include caller functions as part of the benchmark definition, so caller context must be reconstructed from upstream repositories when possible. More importantly, even when real call sites exist, many targets have only sparse caller evidence, as shown in Section ~\ref{sec:DownstreamTasks}. Under such conditions, a subset of tasks can be weakly dependent on the invocation context. Rather than satisfying a rich usage contract, implementing a passing solution for these tasks is often simpler and closer to unconstrained function synthesis. By contrast, \benchmarkname is purpose-built for caller-driven evaluation: it retains naturally occurring calling contexts as a first-class part of each task, and it adopts a curated benchmark construction and validation pipeline to ensure high-quality task specifications and test suites. Consequently, differences in task difficulty and design across benchmarks can lead to gaps in absolute performance.}

{
Beyond benchmark design, the coverage and fidelity of extracted calling contexts can vary with language features and ecosystem patterns, which may in turn affect the strength of the conditioning signal and the gains from invocation-aware training. In statically typed languages, explicit types and compiler-resolved call targets may further enrich invocation contexts and reduce ambiguity. In contrast, runtime dynamism and framework-driven indirections may reduce the coverage of purely static extraction or introduce ambiguity and noise. Common examples include dynamic dispatch, indirect calls via callbacks or function pointers, dynamic loading, and event-driven callbacks.  In such settings, the extraction component may need to incorporate build-aware indexing or lightweight instrumentation to recover faithful invocation relations.
}

\section{Threats to Validity}

\textbf{Threats to Internal Validity.} Threats to internal validity may arise from the implementation of our training and evaluation pipeline. To mitigate potential concerns, we adopt well-established baselines such as CodeT5 and DeepSeek-Coder, using their officially released checkpoints and standardized training configurations. 
For our own models, we ensure reproducibility by fixing random seeds and performing all experiments on a consistent hardware and software stack. Additionally, we verify the correctness of our data processing, function pairing, and static analysis results through multiple rounds of sampling and inspection.

\textbf{Threats to External Validity.} 
Threats to external validity might come from two factors.
First, our benchmark is derived from open-source repositories, which may not fully reflect domain-specific or industrial coding patterns, such as proprietary frameworks, internal APIs, organization-specific conventions, or code that is tightly coupled with build systems and deployment environments. Although we select Python to align with a widely adopted ecosystem and to capture diverse open-source development practices, distribution shifts from open-source to industrial settings may still affect the observed performance and error modes.

Second, our empirical evaluation is limited to a single programming language (Python). Our motivation for using calling contexts is based on a repository-wide property of public code: invocation-aware information (e.g., call sites and surrounding usage patterns) is inherent to codebases, whereas natural language descriptions are often missing or noisy. At the level of the core idea, explicitly utilizing invocation contexts to condition code generation does not rely on Python-specific semantics, since invocation dependencies and usage patterns exist across languages. However, the concrete instantiation of our pipeline, particularly context extraction and call-resolution, depends on language and ecosystem characteristics, such as typing discipline, dynamic dispatch and runtime mechanisms. These factors may affect the coverage and fidelity of extracted contexts, and consequently the effectiveness of invocation-conditioned training. We therefore restrict our empirical claims to Python and leave systematic validation across additional languages and representative framework ecosystems as future work.

Nonetheless, the principle of invocation-conditioned generation is inherently language-agnostic and could be applicable across diverse software ecosystems. We leave the exploration of broader generalization, such as to other programming languages, frameworks, or industrial codebases, as future work.

\section{Conclusion}

In this paper, we address the overlooked yet practical setting of caller-driven code generation. Our work introduces the first invocation-aware framework for code generation. %
Our approach demonstrates that real-world calling contexts provide critical semantic signals for synthesizing behaviorally coherent code. Through \modelname and \benchmarkname, we establish that modeling caller-callee relationships significantly improves generation quality when evaluated under realistic usage scenarios. This highlights that calling contexts are a rich and reliable source of semantic intents, and incorporating them is essential for generating behaviorally coherent code. %

\bibliographystyle{ACM-Reference-Format}
\bibliography{references}

\end{document}